\documentclass[journal]{IEEEtran}
\usepackage{amsmath,amsfonts}
\usepackage{algorithmic}
\usepackage{algorithm}
\usepackage{array}
\usepackage[caption=false]{subfig}
\usepackage{textcomp}
\usepackage{stfloats}
\usepackage{url}
\usepackage{verbatim}
\usepackage{graphicx}
\usepackage{cite}
\usepackage{bm}
\usepackage{booktabs}
\usepackage{hyperref}
\usepackage{multirow, multicol}
\usepackage{makecell}
\usepackage{bbding}
\usepackage{enumitem}
\begin{document}

\title{Towards High-Quality and Efficient Speech Bandwidth Extension with Parallel Amplitude and Phase Prediction}
% <-this % stops a space
\author{Ye-Xin Lu, Yang Ai,~\IEEEmembership{Member,~IEEE}, Hui-Peng Du,  Zhen-Hua Ling,~\IEEEmembership{Senior Member,~IEEE}
% <-this % stops a space
\thanks{The authors are with the National Engineering Research Center of Speech and Language Information Processing, University of Science and Technology of China, Hefei 230027, China (e-mail: yxlu0102@mail.ustc.edu.cn; yangai@ustc.edu.cn; redmist@mail.ustc.edu.cn; zhling@ustc.edu.cn).}
\thanks{This work was funded by the National Nature Science Foundation of China under Grant 62301521, the Anhui Provincial Natural Science Foundation under Grant 2308085QF200, and the Fundamental Research Funds for the Central Universities under Grant WK2100000033.}}



\maketitle

\begin{abstract}
Speech bandwidth extension (BWE) refers to widening the frequency bandwidth range of speech signals, enhancing the speech quality towards brighter and fuller.
This paper proposes a generative adversarial network (GAN) based BWE model with parallel prediction of Amplitude and Phase spectra, named AP-BWE, which achieves both high-quality and efficient wideband speech waveform generation.
The proposed AP-BWE generator is entirely based on convolutional neural networks (CNNs).
It features a dual-stream architecture with mutual interaction, where the amplitude stream and the phase stream communicate with each other and respectively extend the high-frequency components from the source narrowband amplitude and phase spectra.
To improve the naturalness of the extended speech signals, we employ a multi-period discriminator at the waveform level and design a pair of multi-resolution amplitude and phase discriminators at the spectral level, respectively.
Experimental results demonstrate that our proposed AP-BWE achieves state-of-the-art performance in terms of speech quality for BWE tasks targeting sampling rates of both 16 kHz and 48 kHz. 
In terms of generation efficiency, due to the all-convolutional architecture and all-frame-level operations, the proposed AP-BWE can generate 48 kHz waveform samples 292.3 times faster than real-time on a single RTX 4090 GPU and 18.1 times faster than real-time on a single CPU.
Notably, to our knowledge, AP-BWE is the first to achieve the direct extension of the high-frequency phase spectrum, which is beneficial for improving the effectiveness of existing BWE methods.
\end{abstract}


\section{Introduction}
\IEEEPARstart{I}{n} practical speech signal transmission scenarios, limitations in communication devices or transmission channels may lead to the truncation of the frequency bandwidth of speech signals.
The deficiency of high-frequency information can induce distortion, muffling, or a lack of clarity in speech.
Speech bandwidth extension (BWE) aims to supplement the missing high-frequency bandwidth from the low-frequency components, thereby enhancing the quality and intelligibility of the narrowband speech signals.
In the earlier years, the bandwidth of communication devices was extremely limited.
For instance, the bandwidth of speech signals in the public switching telephone network (PSTN) is less than 4 kHz.
So early BWE efforts were primarily focused on extending the bandwidth to a maximum target frequency of 8 kHz.
With the advancement of communication technology, the signal bandwidth that communication devices can transmit has been widening.
Therefore, recent speech BWE research has increasingly focused on extending the bandwidth to the perceptual frequency limits of the human ear (e.g., 22.05 kHz or 24 kHz), enabling applications in high-quality mobile communication, audio remastering and enhancement, and more.
Speech BWE can be applied to various speech signal processing areas, such as text-to-speech (TTS) synthesis \cite{nakamura2014mel}, automatic speech recognition (ASR) \cite{goodarzi2012feature, albahri2016artificial}, speech enhancement (SE) \cite{chennoukh2001speech, mustiere2010bandwidth}, and speech codec \cite{xiao2023multi}.

In the time domain, speech BWE is a conditionally stringent form of speech super-resolution (SR).
Speech SR aims to increase the temporal resolution of low-resolution speech signals by generating high-frequency components, whereas low-resolution speech signals may contain aliased high-frequency components.
In contrast, in BWE, only the low-frequency components are preserved in the narrowband signals.
Consequently, the BWE task poses greater challenges than SR.
Nevertheless, the majority of SR methods are applicable to the BWE task.

Early research on BWE was predominantly based on signal processing techniques, encompassing approaches such as source-filter-based methods \cite{makhoul1979high, chennoukh2001speech}, mapping-based methods \cite{carl1994bandwidth, sadasivan2016joint, unno2005robust}, statistical methods  \cite{pulakka2011speech, ohtani2014gmm, wang2015speech, chen2004hmm, bauer2008hmm, song2009study, yong2014bandwidth}, and so forth.
Source-filter-based methods introduced the source-filter model to extend bandwidth by separately restoring high-frequency residual signals and spectral envelopes.
The high-frequency residual signals are often derived by folding the spectrum of narrowband signals, while predicting high-frequency spectral envelopes presents more challenges.
Mapping-based methods utilized codebook mapping or linear mapping to map lower-band speech representations to their corresponding upper-band envelopes. 
Additionally, statistical methods leveraged Gaussian mixture models (GMMs) and hidden Markov models (HMMs) to establish the mapping relationship between low-frequency spectral parameters and their corresponding high-frequency counterparts. 
Despite the effective performance achieved by these statistical methods in speech BWE, the limited modeling capability of GMMs and HMMs may lead to generating over-smoothed spectral parameters \cite{ling2015deep}.

With the renaissance of deep learning, deep neural networks (DNNs) have shown strong modeling capability.
DNN-based BWE methods can be broadly classified into two categories: waveform-based methods and spectrum-based methods.
In the waveform-based methods, neural networks were employed to learn the direct mapping from the narrowband waveforms to the wideband ones \cite{kuleshov2017audio, ling2018waveform, birnbaum2019temporal, rakotonirina2021self, wang2021towards}, in which both the amplitude and phase information were implicitly restored.
Nevertheless, due to the all-sample-level operations, this category of methods still suffered from the bottleneck of low generation efficiency, especially in generating high-resolution waveforms, limiting the application of this category of methods in low computational power scenarios.
In the spectrum-based methods, neural networks have been adopted to predict high-frequency amplitude-related spectral parameters.
However, it's difficult to parameterize and predict the phase due to its wrapping characteristic and non-structured nature.
The common practice was to replicate \cite{abel2018simple} or mirror-inverse \cite{li2015deep, liu2015novel, gu2016speech} the low-frequency phase to obtain the high-frequency one, which constrained the quality of the extended wideband speech.
Another approach was to use vocoders for phase recovery from the vocal-tract filter parameters \cite{botinhao2006frequency, kontio2007neural, pulakka2011bandwidth} or mel-spectrogram \cite{liu2022neural}.
These vocoder-based methods involved a two-step generation process, where the prediction errors accumulated and the generation efficiency was significantly constrained.
Other methods chose to implicitly recover phase information by predicting the phase-contained spectra, e.g., short-time Fourier transform (STFT) complex spectrum \cite{mandel2023aero} and modified discrete cosine transform (MDCT) spectrum \cite{shuai2023mdct}, but they were still limited in the precise modeling and optimization of phase.
Overall, existing BWE methods have yet to achieve a precise extension of the high-frequency phase, leaving room for improvement in both speech quality and generation efficiency.

In our previous works \cite{ai2023neural, ai2023long}, we proposed a neural speech phase prediction method based on parallel estimation architecture and anti-wrapping losses.
The proposed phase prediction method has been proven to be applicable to various speech-generation tasks, such as speech synthesis \cite{ai2023apnet} and speech enhancement \cite{lu2023mp}.
We also have tried to apply it to speech BWE by predicting the wideband phase spectra from the extended log-amplitude spectra, and the final extended waveforms were obtained through inverse STFT (iSTFT).
However, in our preliminary experiments, we found that this method still faced the same issue of error accumulation and two-step generation as vocoder-based methods, and the low-frequency phase information was not utilized.
Therefore, integrating phase prediction into end-to-end speech BWE might be a preferable option.

Hence, in this paper, we propose AP-BWE, a generative adversarial network (GAN) based end-to-end speech BWE model that achieves high-quality and efficient speech BWE with the parallel extension of amplitude and phase spectra. 
The generator features a dual-stream architecture, with each stream incorporating ConvNeXt \cite{liu2022convnet} as its foundational backbone.
With narrowband log-amplitude and phase spectra as input conditions respectively, the amplitude stream predicts the residual high-frequency log-amplitude spectrum, while the phase stream directly predicts the wrapped wideband phase spectrum.
Additionally, connections are established between these two streams which has been proven to be crucial for phase prediction \cite{yin2020phasen}.
To further enhance the subjective perceptual quality of the extended speech, we first employ the multi-period discriminator (MPD) \cite{kong2020hifi} at the waveform level.
Subsequently, inspired by the multi-resolution discriminator proposed by Jang \emph{et al.} \cite{jang2021univ} to alleviate the spectral over-smoothing, we respectively design a multi-resolution amplitude discriminator (MRAD) and a multi-resolution phase discriminator (MRPD) at the spectral level, aiming to enforce the generator to produce more realistic amplitude and phase spectra.
Experimental results demonstrate that our proposed AP-BWE surpasses state-of-the-art (SOTA) BWE methods in terms of speech quality for target sampling rates of both 16 kHz and 48 kHz.
It's worth noting that while ensuring high generation quality, our model exhibits significantly faster-than-real-time generation efficiency.
For waveform generation at a sampling rate of 48 kHz, our model achieves a generation speed of up to 292.3 times real-time on a single RTX 4090 GPU and 18.1 times real-time on a single CPU.
Compared to the SOTA speech BWE methods, we can also achieve at least a fourfold acceleration on both GPU and CPU.

The main contributions of this work are twofold.
On the one hand, we propose to achieve speech BWE with parallel modeling and optimization of amplitude and phase spectra, which effectively avoids the amplitude-phase compensation issues present in previous works, significantly enhancing the quality of the extended speech.
Additionally, benefiting from the parallel phase estimation architecture and anti-wrapping phase losses, we achieve the precise prediction of the wideband phase spectrum. 
Through the multi-resolution discrimination on the phase spectra, we further enhance the realism of the extended phase at multiple resolutions.
To the best of our knowledge, we are the first to achieve the direct extension of the phase spectrum.
On the other hand, with the all-convolutional architecture and all-frame-level operations, our approach achieves a win-win situation in terms of both generation quality and efficiency.

The rest of this paper is organized as follows.
Section II briefly reviews previous waveform-based and spectrum-based BWE methods.
In Section III, we give details of our proposed AP-BWE framework.
The experimental setup is presented in Section IV, while Section V gives the results and analysis. 
Finally, we give conclusions in Section VI.

\begin{figure*}[htbp!]
  \centering
  \includegraphics[width=\linewidth]{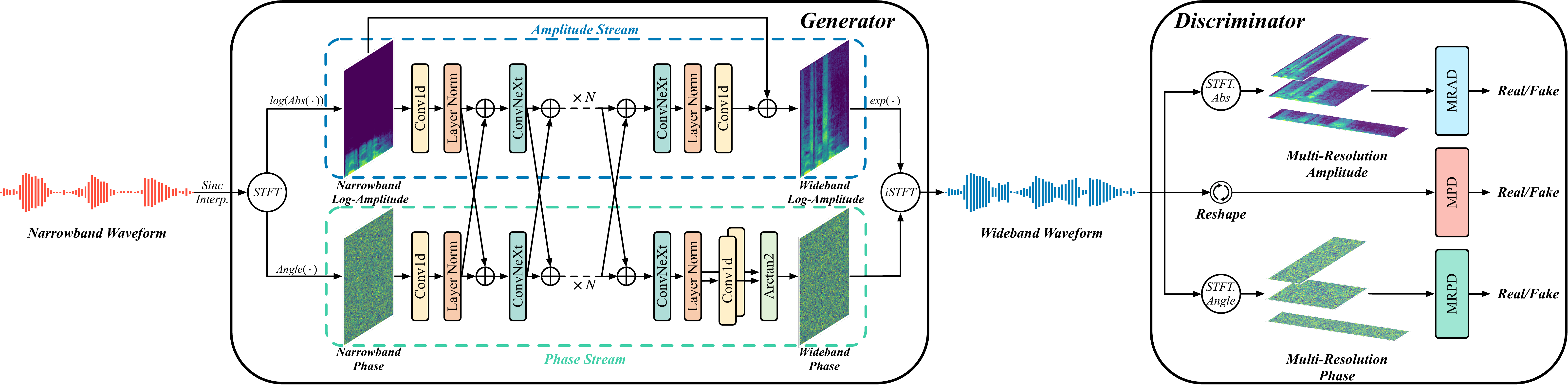}
  \caption{The overall structure of the proposed AP-BWE. The $\mathrm{Abs}(\cdot)$ and $\mathrm{Angle}(\cdot)$ denote the amplitude and phase calculation functions, while $\log(\cdot)$ and $\exp(\cdot)$ denote the logarithmic and exponential functions, respectively. The $\mathrm{Arctan2}$ refers to the two-argument arc-tangent function.}
  \label{fig: Model}
\end{figure*}

\section{Related Work}
\subsection{Waveform-based BWE Methods}
Waveform-based BWE methods aim to directly predict wideband waveforms from narrowband ones without any frequency domain transformation.
AudioUNet \cite{kuleshov2017audio} proposed to use a U-Net \cite{ronneberger2015u} based architecture to reconstruct wideband waveforms without involving specialized audio processing techniques.
TFiLM \cite{birnbaum2019temporal} and AFiLM \cite{rakotonirina2021self} proposed to use recurrent neural networks (RNNs) and the self-attention mechanism \cite{vaswani2017attention} to capture the long-term dependencies, respectively.
Wang \emph{et al.} \cite{wang2021towards} proposed to use an autoencoder convolutional neural network (AECNN) based architecture and cross-domain losses to predict and optimize the wideband waveforms, respectively.
However, the operations in the aforementioned methods were all performed at the sample-point level, leading to relatively lower generation efficiency when compared to spectrum-based methods with frame-level operations.

Recently, diffusion probabilistic models \cite{sohl2015deep, ho2020denoising} have been successfully applied to audio processing tasks.
They have been effectively utilized in speech BWE \cite{lee2021nu, han2022nu, yu2023conditioning} by conditioning the network of the noise predictor with narrowband waveforms, with remarkably high perceptual quality.
The diffusion-based methods decomposed the BWE process into two sub-processes: the forward process, and the reverse process.
In the forward process, Gaussian noises were incrementally added to the narrowband waveforms to obtain whitened latent variables.
Conversely, the wideband waveforms were gradually recovered by removing Gaussian noises step by step in the reverse process.
While these diffusion-based BWE methods have demonstrated promising performance, they still required numerous time steps in the reverse process for waveform reconstruction, thereby imposing significant constraints on generation efficiency.
The comparison between our proposed AP-BWE and these diffusion-based methods will be presented in Section V-B.

\subsection{Spectrum-based BWE Methods}
Spectrum-based BWE methods aim to restore high-frequency spectral parameters for reconstructing wideband waveforms. 
However, as these spectral parameters were mostly amplitude-related, recovering high-frequency phase information remains the primary challenge.
The most primitive method involved replicating or mirror-inversing the low-frequency phase, but such an approach introduces significant errors.
Another method entailed the use of a vocoder to recover the phase from the extended amplitude-related spectrum.
For instance, NVSR \cite{liu2022neural} divided the BWE process into two stages: 1) wideband mel-spectrogram prediction stage; 2) vocoder-based waveform synthesis and post-processing stage.
Initially, NVSR employed ResUNet \cite{diakogiannis2020resunet} to predict wideband mel-spectrograms from narrowband ones. 
Subsequently, these predicted mel-spectrograms were fed into a neural vocoder to reconstruct high-resolution waveforms. 
Finally, the low-frequency components of the high-resolution waveforms were replaced with the original low-frequency ones.

Other methods involved recovering phase information from the phase-contained spectrum. 
AERO \cite{mandel2023aero} directly predicted the wideband short-time complex spectrum from the narrowband one, implicitly recovering both amplitude and phase.
However, the lack of an explicit optimization method for the phase can lead to the compensation effect \cite{wang2021compensation} between amplitude and phase, thereby impacting the quality of generated waveforms.
mdctGAN \cite{shuai2023mdct} utilized the MDCT to encode both amplitude and phase information to a real-valued MDCT spectrum.
While successfully avoiding additional phase prediction through the prediction of the wideband MDCT spectrum, the performance of the MDCT spectrum in waveform generation tasks has been demonstrated to be significantly weaker than that of the STFT spectrum \cite{siuzdak2023vocos}, which may be attributed to the advantageous impact of an over-complete Fourier basis on enhancing training stability \cite{gritsenko2020spectral}.

Both waveform-based and spectrum-based methods mentioned above failed to achieve precise recovery of the high-frequency phase, thereby inevitably limiting the quality of the extended speech.
Building upon our previous work on phase prediction \cite{ai2023neural}, we preliminarily tried to apply it to the BWE task by predicting the wideband phase spectrum from the extended log-amplitude spectrum.
However, we found that this two-stage prediction approach failed to fully leverage the low-frequency phase information in narrowband waveforms, and its prediction errors accumulate across stages. 
Therefore, in this study, we opted to integrate the phase prediction method into the end-to-end speech BWE.

\section{Methodology}
The overview of the proposed AP-BWE is illustrated in Fig.~\ref{fig: Model}.
Given the narrowband waveform $\bm{x} \in \mathbb{R}^{L}$ as input, AP-BWE aims to extend its bandwidth in the spectral domain as well as increase its resolution in the time domain to predict the wideband waveform $\bm{y} \in \mathbb{R}^{nL}$.
Here, $n$ refers to the sampling rate ratio between wideband and narrowband waveforms (i.e., extension factor), while $nL$ and $L$ represent the length of the wideband and narrowband waveforms, respectively.
Specifically, the narrowband waveform $\bm{x}$ is first interpolated $n$ times using the sinc filter to match the temporal resolution of $\bm{y}$.
Subsequently, the narrowband amplitude spectrum $\bm{X}_a \in \mathbb{R}^{T\times F}$ and wrapped phase spectrum $\bm{X}_p \in \mathbb{R}^{T\times F}$ are extracted from the interpolated narrowband waveform through STFT, where $T$ and $F$ denote the number of temporal frames and frequency bins, respectively.
Through the mutual coupling of the amplitude stream and the phase stream, AP-BWE predicts wideband log-amplitude spectrum $\log(\bm{\hat{Y}}_a) \in \mathbb{R}^{T\times F}$ as well as wideband wrapped phase spectrum $\bm{\hat{Y}}_p \in \mathbb{R}^{T\times F}$ separately from $\log(\bm{X}_a)$ and $\bm{X}_p$.
Eventually the wideband waveform $\bm{\hat{y}} \in \mathbb{R}^{nL}$ was reconstructed through iSTFT.
The details of the model structure and training criteria are described as follows.

\subsection{Model Structure}
\subsubsection{Generator}
\begin{figure}[t!]
  \centering
  \includegraphics[width=\linewidth]{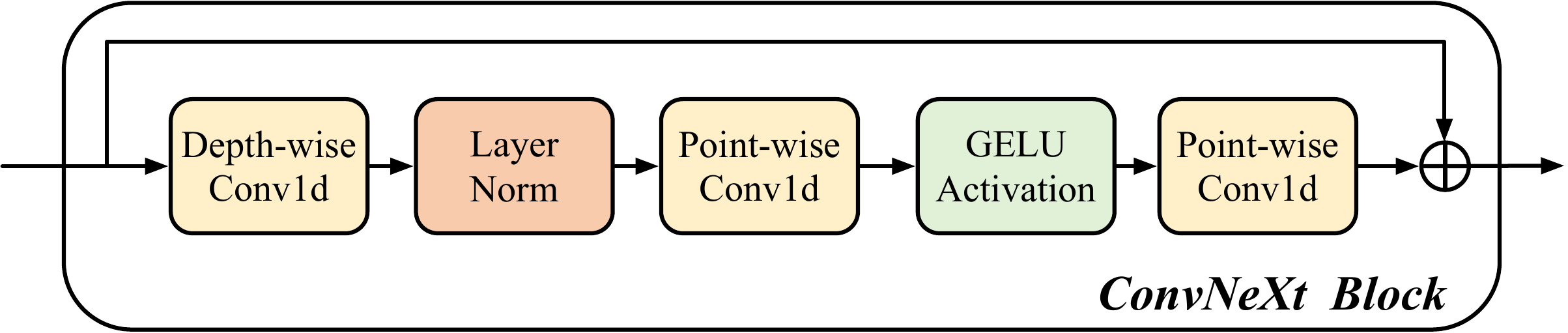}
  \caption{Details of the ConvNeXt block \cite{liu2022convnet}: Each ConvNeXt block consists of a $7 \times 1$ depth-wise convolution, followed by layer normalization, a $1 \times 1$ point-wise convolution for dimensionality projection with an expansion factor of 3, a GELU activation layer, and another $1 \times 1$ point-wise convolution for dimensionality restoration followed by residual connection.}
  \label{fig: ConvNeXt}
\end{figure}

We denote the generator of our proposed AP-BWE as $G$, and $\bm{\hat{y}} = G(\bm{x})$.
As depicted in Fig.~\ref{fig: Model}, the generator $G$ comprises a dual-stream architecture, which is entirely based on convolutional neural networks.
Both the amplitude and phase streams utilize the ConvNeXt \cite{liu2022convnet} as the foundational backbone due to its strong modeling capability.
The original two-dimensional convolution-based ConvNeXt is modified into a one-dimensional convolution-based version and integrated it into our model.
As depicted in Fig.~\ref{fig: ConvNeXt}, the ConvNeXt block is a cascade of a large-kernel-sized depth-wise convolutional layer and a pair of point-wise convolutional layers that respectively expand and restore feature dimensions.
Layer normalization \cite{ba2016layer} and Gaussian error linear unit (GELU) activation \cite{hendrycks2017gaussian} are interleaved between the layers.
Finally, the residual connection is added before the output to prevent the gradient from vanishing.

The amplitude stream comprises a convolutional layer, $N$ ConvNeXt clocks, and another convolutional layer, with the aim to predict the residual high-frequency log-amplitude spectrum and add it to the narrowband $\log(\bm{X}_a)$ to obtain the wideband log-amplitude spectrum $\log(\bm{\hat{Y}}_a)$.
Differs slightly from the amplitude stream, the phase stream incorporates two output convolutional layers to respectively predict the pseudo-real part component $\bm{\hat{Y}}_p^{(r)}$ and pseudo-imaginary part component $\bm{\hat{Y}}_p^{(i)}$, and further calculate the wrapped phase spectrum $\bm{\hat{Y}}_p$ from them with the two-argument arc-tangent (Arctan2) function:
\begin{equation}
	\bm{\hat{Y}}_p = \arctan\bigg(\frac{\bm{\hat{Y}}_p^{(i)}}{\bm{\hat{Y}}_p^{(r)}}\bigg) - \frac{\pi}{2} \cdot \mathrm{Sgn}^*(\bm{\hat{Y}}_p^{(i)}) \cdot [\mathrm{Sgn}^*(\bm{\hat{Y}}_p^{(r)}) - 1],
\end{equation}
where $\mathrm{arctan}(\cdot)$ denotes the arc-tangent function, and $\mathrm{Sgn}^*(x)$ is a redefined symbolic function: when $x \geq 0$, $\mathrm{Sgn}^*(x) = 1$, otherwise $\mathrm{Sgn}^*(x) = -1$.
% The direct prediction of the wrapped phase in this manner has been proven effective in our previous work \cite{ai2023neural}.
Additionally, connections are established between two streams for information exchange, which is crucial for phase prediction \cite{yin2020phasen}.
Finally, the predicted wideband waveform $\bm{\hat{y}} \in \mathbb{R}^{nL}$ is reconstructed from $\bm{\hat{Y}}_a$ and $\bm{\hat{Y}}_p$ using iSTFT:
\begin{equation}
    \begin{aligned}
      	\bm{\hat{y}} &= \mathrm{iSTFT}(\bm{\hat{Y}}_a \cdot e^{j\bm{\hat{Y}}_p}) \\
                     &= \mathrm{iSTFT}(\bm{\hat{Y}}_r + j\bm{\hat{Y}}_i),
    \end{aligned}
\end{equation}
where $\bm{\hat{Y}}_r = \bm{\hat{Y}}_a \cdot \cos(\bm{\hat{Y}}_p) \in \mathbb{R}^{T\times F}$ and $\bm{\hat{Y}}_i = \bm{\hat{Y}}_a \cdot \sin(\bm{\hat{Y}}_p) \in \mathbb{R}^{T\times F}$ denote the real and imaginary parts of the extended short-time complex spectrum, respectively.

\begin{figure}[t!]
  \centering
  \includegraphics[width=\linewidth]{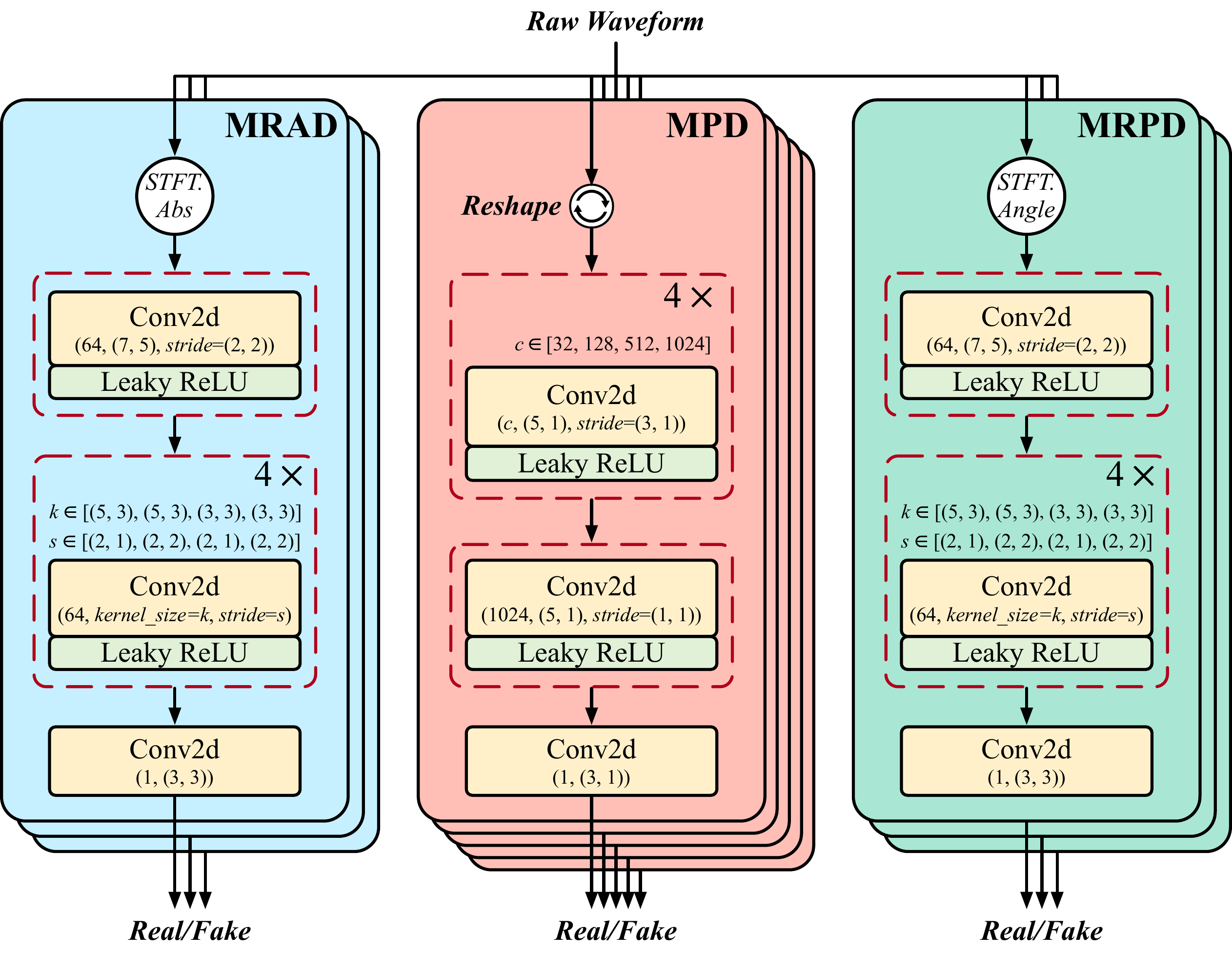}
  \caption{Details of the discriminators. The parameters inside the parentheses for each convolutional layer respectively represent the number of channels, kernel size, and stride.}
  \label{fig: Disc}
\end{figure}
\subsubsection{Discriminator}

Directly predicting amplitude and phase and then reconstructing the speech waveform through iSTFT can result in over-smoothed spectral parameters, manifesting as a robotic or muffled quality in the reconstructed waveforms.
To this end, we utilize discriminators defined both in the spectral domain and time domain to guide generator $G$ in generating spectra and waveforms that closely resemble real ones.
Firstly, considering that the speech signal is composed of sinusoidal signals with various frequencies, with some frequency bands generated through BWE.
Due to the statistical characteristics of speech signals varying in different frequency bands, we employ an MPD \cite{kong2020hifi} to capture periodic patterns, with the aim of matching the natural wideband speech across multiple frequency bands.
Moreover, since the statistical characteristics of amplitude and phase also differ across frequency bands, and the sole utilization of  MPD cannot cover all frequency bands, we consequently define discriminators on both amplitude and phase spectra.
Drawing inspiration from the multi-resolution discriminator \cite{jang2021univ}, we respectively introduce MRAD and MRPD, with the aim to capture full-band amplitude and phase patterns at various resolutions. 
The details of MPD, MRAD, and MRPD are described as follows.
\begin{itemize}[left=0em]
\item {}{\textbf{Multi-Period Discriminator}}:
As depicted in Fig.~\ref{fig: Disc}, the MPD contains multiple sub-discriminators, each of which comprises a waveform two-dimensional reshaping module, multiple convolutional layers with an increasing number of channels, and an output convolutional layer.
Firstly, the reshaping module reshapes the one-dimensional raw waveform into a two-dimensional format by sampling with a period $p$, which is set to prime numbers to prevent overlaps.
Subsequently, the reshaped waveform undergoes multiple convolutional layers with leaky rectified linear unit (ReLU) activation \cite{maas2013rectifier} before finally producing the discriminative score, which indicates the likelihood that the input data is real.
\item {}{\textbf{Multi-Resolution Discriminators}}:
As depicted in Fig.~\ref{fig: Disc}, both MRAD and MRPD share a unified structure.
They both consist of multiple sub-discriminators, each comprising a spectrum extraction module and multiple convolutional layers interleaved with leaky ReLU activation to capture features along both temporal and frequency axes.
The raw waveform first undergoes an initial transformation into amplitude or phase spectra using STFT with diverse parameter sets, encompassing FFT point number, window size, and hop size. 
Subsequently, the multi-resolution amplitude or phase spectra are processed through multiple convolutional layers to yield the discriminative score.
\end{itemize}

\subsection{Training Criteria}
\subsubsection{Spectrum-based Losses}
We first define loss functions in the spectral domain to capture time-frequency distributions and generate realistic spectra.

\begin{itemize}[left=0em]
\item {}{\textbf{Amplitude Spectrum Loss}}:
The amplitude spectrum loss is the mean square error (MSE) of the wideband log-amplitude spectrum $\log(\bm{Y}_a) \in \mathbb{R}^{T\times F}$ and the extended log-amplitude spectrum $\log(\bm{\hat{Y}}_a)$, which is defined as: 
\begin{equation}
	\mathcal{L}_{A} = \frac{1}{TF}\mathbb{E}_{(\bm{Y}_a, \bm{\hat{Y}}_a)} \bigg[\Vert \log\big(\frac{\bm{Y}_a}{\bm{\hat{Y}}_a}\big) \Vert_\mathrm{F}^2\bigg].
\end{equation}

\item {}{\textbf{Phase Spectrum Loss}}:
Considering the phase wrapping issue, we follow our previous work \cite{ai2023neural} to use three anti-wrapping losses to explicitly optimize the wrapped phase spectrum, which are respectively defined as the mean absolute error (MAE) between the anti-wrapped wideband and extended instantaneous phase (IP) spectra $\bm{Y}_p$ and $\bm{\hat{Y}}_p$,  group delay (GD) spectra $\bm{Y}_{GD}$ and $\bm{\hat{Y}}_{GD}$, and instantaneous angular frequency (IAF) spectra $\bm{Y}_{IAF}$ and $\bm{\hat{Y}}_{IAF}$:
\begin{equation}
\label{eq: phase_ip}
    \mathcal{L}_{IP} = \frac{1}{TF}\mathbb{E}_{(\bm{Y}_p, \bm{\hat{Y}}_p)} \bigg[\Vert f_{AW}(\bm{Y}_p - \bm{\hat{Y}}_p) \Vert_1 \bigg],
\end{equation}
\begin{equation}
\label{eq: phase_gd}
    \mathcal{L}_{GD} = \frac{1}{TF}\mathbb{E}_{(\bm{Y}_{GD}, \bm{\hat{Y}}_{GD})} \bigg[\Vert f_{AW}(\bm{Y}_{GD} - \bm{\hat{Y}}_{GD})) \Vert_1 \bigg], 
\end{equation}
\begin{equation}
\label{eq: phase_iaf}
    \mathcal{L}_{IAF} = \frac{1}{TF}\mathbb{E}_{(\bm{Y}_{IAF}, \bm{\hat{Y}}_{IAF})} \bigg[\Vert f_{AW}(\bm{Y}_{IAF} - \bm{\hat{Y}}_{IAF})) \Vert_1 \bigg],
\end{equation}
where $(\bm{Y}_{GD}, \bm{\hat{Y}}_{GD}) = (\Delta_{DF}\bm{Y}_p, \Delta_{DF}\bm{\hat{Y}_p})$ and  $(\bm{Y}_{IAF}, \bm{\hat{Y}}_{IAF}) = (\Delta_{DT}\bm{Y}_p, \Delta_{DT}\bm{\hat{Y}_p})$.
The $\Delta_{DF}$ and $\Delta_{DT}$ represent the differential operator along the frequency and temporal axes, respectively.
The $f_{AW}(x)$ denotes the anti-wrapping function, which is defined as $f_{AW}(x) = \vert x - 2\pi \cdot {\rm round} \big(\frac{x}{2\pi}\big) \vert, x \in \mathbb{R}$.
The final phase spectrum loss is the sum of these three anti-wrapping losses:
\begin{equation}
	\mathcal{L}_{P} = \mathcal{L}_{IP} + \mathcal{L}_{GD} + \mathcal{L}_{IAF}.
\end{equation}

\item {}{\textbf{Complex Spectrum Loss}}:
To further optimize the amplitude and phase within the complex spectrum and enhance the spectral consistency of iSTFT, we define the MSE loss between the wideband short-time complex spectrum $(\bm{Y}_r, \bm{Y}_i)  \in \mathbb{R}^{T \times F \times 2}$ and extend short-time complex spectrum $(\bm{\hat{Y}}_r, \bm{\hat{Y}}_i)  \in \mathbb{R}^{T \times F \times 2}$ as well as the MSE loss between $(\bm{\hat{Y}}_r, \bm{\hat{Y}}_i)$ and re-extracted short-time complex spectrum $(\bm{\hat{Y}}_r', \bm{\hat{Y}}_i') \in \mathbb{R}^{T \times F \times 2}$ from the extended waveform $\bm{\hat{y}}$.
So the complex spectrum loss is defined as:
\begin{equation}
    \begin{aligned}
        \mathcal{L}_{C} =& \frac{1}{TF} \mathbb{E}_{(\bm{Y}_r,\bm{Y}_i), (\bm{\hat{Y}}_r, \bm{\hat{Y}}_i)} \bigg[\Vert (\bm{Y}_r,\bm{Y}_i) - (\bm{\hat{Y}}_r, \bm{\hat{Y}}_i) \Vert_\mathrm{F}^2\bigg] \\
        +& \frac{1}{TF} \mathbb{E}_{(\bm{\hat{Y}}_r, \bm{\hat{Y}}_i), (\bm{\hat{Y}}_r', \bm{\hat{Y}}_i')} 
        \bigg[\Vert (\bm{\hat{Y}}_r, \bm{\hat{Y}}_i) - (\bm{\hat{Y}}_r', \bm{\hat{Y}}_i') \Vert_\mathrm{F}^2\bigg].
    \end{aligned} 
\end{equation}

\item {}{\textbf{Final Spectral Loss}}:
The final spectral loss is the linear combination of the spectrum-based losses mentioned above:
\begin{equation}
    \mathcal{L}_{S} = \lambda_{A}\mathcal{L}_{A} + \lambda_{P}\mathcal{L}_{P} + \lambda_{C}\mathcal{L}_{C},
\end{equation}
where $\lambda_{A}$, $\lambda_{P}$, and $\lambda_{C}$ are hyper-parameters and we set them to 45, 100, and 45, respectively.
\end{itemize}

\subsubsection{GAN-based Losses}
\begin{itemize}[left=0em]
\item {}{\textbf{GAN Loss}}:
For brevity, we represent MPD, MRAD, and MRPD collectively as $D$.
The discriminator $D$ and generator $G$ are trained alternately. 
The discriminator is trained to classify wideband samples as 1 and samples extended by the generator as 0; conversely, the generator is trained to generate samples that approach being classified as 1 by the discriminator as closely as possible.
We use the hinge GAN loss \cite{zeghidour2021soundstream} which is defined as:
\begin{equation}
    \setlength{\belowdisplayskip}{3pt}
    \begin{aligned}
    \mathcal{L}_{adv}(D;G) =& \mathbb{E}_{\bm{x}} \bigg[\max(0, 1+D(G(\bm{x})) \bigg] \\
                           +& \mathbb{E}_{\bm{y}} \bigg[ \max(0, 1-D(\bm{y})) \bigg],
    \end{aligned}
\end{equation}
\begin{equation}
    \setlength{\abovedisplayskip}{3pt}
    \mathcal{L}_{adv}(G;D) = \mathbb{E}_{\bm{x}} \bigg[\max(0, 1-D(G(\bm{x}))) \bigg].
\end{equation}

\item {}{\textbf{Feature Matching Loss}}:
To encourage the generator to produce samples that not only fool the discriminator but also match the features of real samples at multiple levels of abstraction, we define the feature matching loss \cite{kumar2019melgan} between the features extracted from the natural wideband waveforms and those from the extended waveforms at certain intermediate layers of the discriminator as follows:
\begin{equation}
    \mathcal{L}_{FM}(G;D) = \mathbb{E}_{(\bm{x}, \bm{y})} \bigg[ \sum_{i=1}^{M} \frac{1}{N_i} \Vert D^i(\bm{y})-D^i(G(\bm{x})) \Vert_1 \bigg],
\end{equation}
where $M$ denotes the number of layers in the discriminator, $D^i$ and $N_i$ denotes the features and the number of features in the $i$-th layer of the discriminator, respectively.

\end{itemize}

\subsubsection{Final Loss}
Since the discriminator $D$ is a set of sub-discriminators of MPD, MRAD, and MRPD, the final losses of the generator and discriminator are defined as:
\begin{gather}
    \setlength{\belowdisplayskip}{3pt}
    \mathcal{L}_{G} = \sum_{k=1}^{K} \bigg[\lambda_{adv} \mathcal{L}_{adv}(G;D_k) +\lambda_{FM} \mathcal{L}_{FM}(G;D_k) \bigg] + \lambda_{S} \mathcal{L}_{S}, \\
    \setlength{\abovedisplayskip}{3pt}
    \mathcal{L}_{D} = \sum_{k=1}^{K} \mathcal{L}_{adv}(D_k; G).
\end{gather}
where $K$ denotes the numbers of sub-discriminators, and $D_k$ denotes the $k$-th sub-discriminator in MPD, MRAD, and MRPD.
$\lambda_{adv}$, $\lambda_{FM}$, and $\lambda_{S}$ are hyper-parameters and in all our experiments, we set $\lambda_{S}=1$. For MPD, we set $\lambda_{adv} = 1$ and $\lambda_{FM} = 1$, while for MRAD and MRPD, we set $\lambda_{adv} = 0.1$ and $\lambda_{FM} = 0.1$.
\section{Experimental Setup}

\subsection{Data Configuration}
We trained all models on the VCTK-0.92 dataset \cite{yamagishi2019cstr}, which contains approximately 44 hours of speech recordings from 110 speakers with diverse accents. 
Adhering to the data preparation approach adopted in previous speech BWE studies \cite{liu2022neural, lee2021nu, han2022nu, yu2023conditioning, mandel2023aero}, we exclusively utilized the $mic1$-microphone data and excluded speakers $p280$ and $p315$ due to technical issues. Among the remaining 108 speakers, the last 8 were allocated for testing, while the remaining 100 were used for training.
Given the historical focus of early speech BWE methods on a sampling rate of 16 kHz and the contemporary emphasis on higher target sampling rates (e.g., 44.1 kHz and 48 kHz) in recent methods, we employed the original VCTK-0.92 dataset with a 48 kHz sampling rate for high-sampling-rate BWE experiments. 
Subsequently, we downsampled the VCTK-0.92 dataset to 16 kHz for low-sampling-rate BWE experiments.

To generate pairs of wideband and narrowband speech signals, we employed a sinc filter to eliminate high-frequency components in the speech signals above a specified bandwidth. This process retained only the low-frequency components, ensuring no aliasing occurred.
For experiments targeting a 16 kHz sampling rate, we configured the downsampling rate $n$ to 2, 4, and 8, corresponding to the extension from 8 kHz, 4 kHz, and 2 kHz to 16 kHz, respectively. 
In experiments aiming for a 48 kHz sampling rate, we set the downsampling rate $n$ to 2, 3, 4, and 6, denoting the extension from 24 kHz, 16 kHz, 12 kHz, and 8 kHz to 48 kHz, respectively. 

\subsection{Model Details}
We used the same configuration for experiments with target sampling rates of 16 kHz and 48 kHz.
For training our proposed AP-BWE model, all the audio clips underwent silence trimming with VCTK silence labels \footnote{\href{https://github.com/nii-yamagishilab/vctk-silence-labels}{https://github.com/nii-yamagishilab/vctk-silence-labels}.} and sliced into 8000-sample-point segments.
To extract the amplitude and phase spectra from raw waveforms, we used STFT with the FFT point number of 1024, Hanning window size, and hop size of 320 and 80 sample points, respectively.
So for the training set, the number of frequency bins $F$ is 513, and the number of temporal frames $T$ is 101.

For the generator, the number of the ConvNeXt block $N$ was set to 8.
The period $p$ for each sub-discriminator in the MPD was configured as 2, 3, 5, 7, and 11. 
In the case of MRAD and MRPD, the FFT point numbers, rectangular window sizes, and hop sizes of the STFT parameter sets were set to [512, 128, 512], [1024, 256, 1024], and [2048, 512, 2048] for the three sub-discriminators, respectively.
Both the generator and discriminator were trained until 500k steps using the AdamW optimizer \cite{loshchilov2017decoupled}, with $\beta_1=0.8$, $\beta_2=0.99$, and weight decay $\lambda=0.01$.
The learning rate was set initially to $2\times 10^{-4}$ and scheduled to decay with a factor of 0.999 at every epoch.
\footnote{Source codes and audio samples of the proposed AP-BWE can be accessed at \href{https://github.com/yxlu-0102/AP-BWE}{https://github.com/yxlu-0102/AP-BWE}.}

\subsection{Evaluation Metrics}
\subsubsection{Metrics on Speech Quality}
We comprehensively evaluated the quality of the extended speech signals using metrics defined on the amplitude spectra, phase spectra, and reconstructed speech waveforms, including:
\begin{itemize}[left=0em]
\item {}{\textbf{Log-Spectral Distance (LSD)}}:
LSD is a commonly used objective metric in the BWE task.
Given the wideband and extended speech waveform $\bm{y}$ and $\bm{\hat{y}}$, their corresponding amplitude spectra $\bm{Y}_a \in \mathbb{R}^{T\times F}$ and $\bm{\hat{Y}}_a \in \mathbb{R}^{T\times F}$ were first extracted using STFT with the FFT point number of 2048, Hanning window size of 2048, and hop size of 512.
Then the LSD is defined as:
\begin{equation}
    \mathrm{LSD} = \frac{1}{T} \sum_{t=1}^{T} \sqrt{\frac{1}{F} \sum_{f=1}^{F} \bigg(\log_{10}\big(\frac{\bm{Y}_a[t, f]}{\bm{\hat{Y}}_a[t, f]}\big)\bigg)^2}
\end{equation}

\item {}{\textbf{Anti-Wrapping Phase Distance (AWPD)}}:
To assess the model's capability of recovering high-frequency phase, on the basis of the anti-wrapping losses defined in Eq.~\ref{eq: phase_ip}-\ref{eq: phase_iaf}, we defined three anti-wrapping phase metrics to evaluate the extended phase instantaneous error as well as its continuity in both the temporal and frequency domains:

\begin{small}
    \begin{gather}
        \mathrm{AWPD}_{IP} = \frac{1}{T} \sum_{t=1}^{T} \sqrt{\frac{1}{F} \sum_{f=1}^{F} f_{AW}^2(\bm{Y}_p[t, f]-\bm{\hat{Y}}_p[t, f])},\\
        \mathrm{AWPD}_{GD} = \frac{1}{T} \sum_{t=1}^{T} \sqrt{\frac{1}{F} \sum_{f=1}^{F} f_{AW}^2(\bm{Y}_{GD}[t, f] - \bm{\hat{Y}}_{GD}[t, f])},\\
        \mathrm{AWPD}_{IAF} = \frac{1}{T} \sum_{t=1}^{T} \sqrt{\frac{1}{F} \sum_{f=1}^{F} f_{AW}^2(\bm{Y}_{IAF}[t, f] - \bm{\hat{Y}}_{IAF}[t, f])},
    \end{gather}
\end{small}

where all the spectra are extracted using the same STFT parameters as those used in LSD.

\item {}{\textbf{Virtual Speech Quality Objective Listener (ViSQOL)}}:
To access the overall perceived audio quality of the extended speech signals in an objective manner, we employed the ViSQOL \footnote{\href{https://github.com/google/visqol}{https://github.com/google/visqol}.} \cite{chinen2020visqol} which uses a spectral-temporal measure of similarity between a reference and a test speech signal to produce a mean opinion score - listening quality objective (MOS-LQO) score.
For the audio mode of ViSQOL at a required sampling rate of 48 kHz, the MOS-LQO score ranges from 1 to 4.75, the higher the better.
For the speech mode of ViSQOL at a required sampling rate of 16 kHz, the MOS-LQO score ranges from 1 to 5.

\item {}{\textbf{Mean Opinion Score (MOS)}}:
To further subjectively access the overall audio quality, MOS tests were conducted to evaluate the naturalness of the wideband speech and speech waveforms extended by the speech BWE models.
Defining the extension ratio as the ratio between the target sampling rate and the source sampling rate, we selected configurations with the highest extension ratios for subjective evaluations.
In each MOS test, twenty utterances from the test set were evaluated by at least 30 native English listeners on the crowd-sourcing platform Amazon Mechanical Turk.
For each utterance, listeners were asked to rate a naturalness score between 1 and 5 with an interval of 0.5.
All the MOS results were reported with 95\% confidence intervals (CI).
We also conducted paired $t$-tests to assess the significance of differences between our proposed AP-BWE and the baseline models, reporting $p$-values to indicate the statistical significance of these comparisons.

\end{itemize}

\subsubsection{Metrics on Generation Efficiency}
We first used the real-time factor (RTF) to evaluate the inference speed of the model.
The RTF is defined as the ratio of the total inference time for processing narrowband source signals into wideband output signals, to the total duration of the wideband signals. 
In our implementation, RTF was calculated using the complete test set on an RTX 4090 GPU and an Intel(R) Xeon(R) Silver 4310 CPU (2.10 GHz). 
Additionally, we used floating point operations (FLOPs) to assess the computational complexity of the model.
All the FLOPs were calculated using 1-second speech signals as inputs to the models.

\subsubsection{Metrics on Speech Intelligibility}
The main frequency components of human speech are concentrated within the range of approximately 300 Hz to 3400 Hz. This frequency range encompasses crucial information for vowels and consonants, significantly impacting speech intelligibility. 
Consequently, we analyzed the intelligibility of waveforms extended by speech BWE methods with the target sampling rate of 16 kHz.
Firstly, we employed an advanced ASR model, Whisper \cite{radford2023robust} to transcribe the extended 16 kHz speech signals into corresponding texts. 
% We used the ``base.en'' version of Whisper in our implementation.
Subsequently, we calculated the word error rate (WER) and character error rate (CER) based on the transcription results.
Additionally, short-time objective intelligibility (STOI) was also included as an objective metric to indicate the percentage of speech signals that are correctly understood.

\begin{table*}[ht!]\scriptsize
  \caption{Experimental Results in Speech Quality (LSD and ViSQOL) and Generation Efficiency (RTF and FLOPs) for BWE Methods Evaluated on the VCTK Dataset with Target Sampling Rate of 16 kHz, Where in RTF ($a\times$) Representing $a$ Times Real-Time}
  \label{tab: results_16k}
  \centering
  \resizebox{\textwidth}{!}{
  \begin{tabular}{l|cc|cc|cc|cc|c}
    \toprule
    \multirow{2}{*}{Method} & \multicolumn{2}{c|}{$\mathrm{8 kHz \rightarrow 16 kHz}$} & \multicolumn{2}{c|}{$\mathrm{4 kHz \rightarrow 16 kHz}$} & \multicolumn{2}{c|}{$\mathrm{2 kHz \rightarrow 16 kHz}$} & \multirow{2}{*}{$\mathrm{RTF (CPU)}$} & \multirow{2}{*}{$\mathrm{RTF (GPU)}$} & \multirow{2}{*}{$\mathrm{FLOPs}$} \\ 
    \cmidrule{2-7}
    & $\mathrm{LSD}$ & $\mathrm{ViSQOL}$ & $\mathrm{LSD}$ & $\mathrm{ViSQOL}$ & $\mathrm{LSD}$ & $\mathrm{ViSQOL}$ &  &  \\
    \midrule
    sinc & 1.80 & 4.34 & 2.68 & 3.52 & 3.15 & 2.73 & - & - & - \\
    \midrule
    TFiLM \cite{birnbaum2019temporal} & 1.31 & 4.46 & 1.65 & 3.84 & 1.97 & 3.10 & 0.3287 (3.04$\times$) & 0.0244 (41.01$\times$) & 232.85G \\
    AFiLM \cite{rakotonirina2021self} & 1.24 & 4.39 & 1.63 & 3.83 & 1.79 & 2.75 & 0.5029 (1.99$\times$) & 0.0477 (20.96$\times$) & 260.76G \\
    NVSR \cite{liu2022neural}         & 0.79 & 4.52 & 0.95 & 4.11 & 1.10 & 3.41 & 0.7577 (1.32$\times$) & 0.0512 (19.54$\times$) & 34.28G \\
    AERO \cite{mandel2023aero}        & 0.87 & 4.57 & 1.00 & 4.19 & - & - & 0.4395 (2.28$\times$) & 0.0217 (46.01$\times$) & 141.77G \\
    \midrule
    AP-BWE*   & 0.71 & 4.66 & 0.88 & 4.28 & \textbf{0.99} & \textbf{3.77} & \multirow{2}{*}{\textbf{0.0338 (29.61$\times$)}} & \multirow{2}{*}{\textbf{0.0026 (382.56$\times$)}} & \multirow{2}{*}{\textbf{5.97G}} \\
    \textbf{AP-BWE} & \textbf{0.69} & \textbf{4.71} & \textbf{0.87} & \textbf{4.30} & \textbf{0.99} & 3.76 & & \\
    \bottomrule
  \end{tabular}}
\end{table*}
\section{Results and Analysis}
\subsection{BWE Experiments Targeting 16 kHz}
\subsubsection{Baseline Methods}
For BWE targeting a 16 kHz sampling rate, we first used the sinc filter interpolation as the lower-bound method, and further compared our proposed AP-BWE with two waveform-based methods (TFiLM \cite{birnbaum2019temporal} and AFiLM \cite{rakotonirina2021self}), a vocoder-based method (NVSR \cite{liu2022neural}), and a complex-spectrum-based method (AERO \cite{mandel2023aero}).
For TFiLM and AFiLM, we used their official implementations \footnote{\href{https://github.com/ncarraz/AFILM}{https://github.com/ncarraz/AFILM}.}.
However, their original papers used the old-version VCTK dataset \cite{veaux2017cstr} and 
employed subsampling to obtain the narrowband waveforms, which aliased high-frequency components.
Thus, they performed not a strict BWE task but an SR task.
For a fair comparison, we re-trained the TFiLM and AFiLM models with our data-preprocessing manner on the VCTK-0.92 dataset until 50 epochs.
For AERO and NVSR, we used their official implementations \footnote{\href{https://github.com/haoheliu/ssr_eval}{https://github.com/haoheliu/ssr\_eval}.} \footnote{\href{https://github.com/slp-rl/aero}{https://github.com/slp-rl/aero}.}.
Notably, AERO did not conduct the experiment at a 2 kHz source sampling rate, and thus, this result was excluded from our analysis.

Additionally, considering some recent BWE methods \cite{liu2022neural, han2022nu, yu2023conditioning} demonstrated the ability to handle various source sampling rates with a single model, we also trained our AP-BWE with the source sampling rate uniformly sampled from 2 kHz to 8 kHz, denoted as AP-BWE*, to unifiedly extend speech signals at all these three sampling rates to 16 kHz.
\begin{table*}[ht!]\large
  \caption{Phase-related Evaluation Results for BWE methods Evaluated on the VCTK Dataset with Target Sampling Rate of 16 kHz}
  \label{tab: results_pha_16k}
  \centering
  \resizebox{0.9\textwidth}{!}{
  \begin{tabular}{l|ccc|ccc|ccc}
    \toprule
    \multirow{2}{*}{Method} & \multicolumn{3}{c|}{$\mathrm{8 kHz \rightarrow 16 kHz}$} & \multicolumn{3}{c|}{$\mathrm{4 kHz \rightarrow 16 kHz}$} & \multicolumn{3}{c}{$\mathrm{2 kHz \rightarrow 16 kHz}$} \\
    \cmidrule{2-10}
    & $\mathrm{AWPD}_{IP}$ & $\mathrm{AWPD}_{GD}$ & $\mathrm{AWPD}_{IAF}$ & $\mathrm{AWPD}_{IP}$ & $\mathrm{AWPD}_{GD}$ & $\mathrm{AWPD}_{IAF} $ & $\mathrm{AWPD}_{IP}$ & $\mathrm{AWPD}_{GD}$ & $\mathrm{AWPD}_{IAF}$ \\
    \midrule
    sinc  & 1.27 & 0.87 & 1.06 & 1.57 & 1.18 & 1.28 & 1.69 & 1.34 & 1.38 \\
    \midrule
    TFiLM \cite{birnbaum2019temporal} & 1.28 & 0.91 & 1.07 & 1.54 & 1.18 & 1.27 & 1.68 & 1.35 & 1.37 \\
    AFiLM \cite{rakotonirina2021self} & 1.32 & 0.98 & 1.11 & 1.54 & 1.19 & 1.27 & 1.70 & 1.38 & 1.39 \\
    NVSR \cite{liu2022neural} & 1.38 & 0.89 & 1.11 & 1.61 & 1.14 & 1.29 & 1.72 & 1.29 & 1.38 \\
    AERO \cite{mandel2023aero} & 1.31 & 0.93 & 1.08 & 1.56 & 1.15 & 1.27 &  -   &  -   &  -   \\
    \midrule
    AP-BWE* & 1.27 & 0.86 & 1.05 & \textbf{1.53} & \textbf{1.12} & \textbf{1.25} & \textbf{1.67} & \textbf{1.27} & \textbf{1.35} \\
    \textbf{AP-BWE} & \textbf{1.26} & \textbf{0.84} & \textbf{1.04} & \textbf{1.53} & \textbf{1.12} & \textbf{1.25} & \textbf{1.67} & \textbf{1.27} & \textbf{1.35} \\
    \bottomrule
  \end{tabular}}
\end{table*}
% \begin{table*}[ht!]\normalsize
%   \caption{Experimental Results in Intelligibility for BWE methods Evaluated on the VCTK Dataset with Target Sampling Rate of 16 kHz}
%   \label{tab: results_asr}
%   \centering
%   % \renewcommand\arraystretch{1.2}
%   \resizebox{0.9\textwidth}{!}{
%   \begin{tabular}{l|ccc|ccc|ccc}
%     \toprule
%     \multirow{2}{*}{Method} & \multicolumn{3}{c|}{$\mathrm{8 kHz \rightarrow 16 kHz}$} & \multicolumn{3}{c|}{$\mathrm{4 kHz \rightarrow 16 kHz}$} & \multicolumn{3}{c}{$\mathrm{2 kHz \rightarrow 16 kHz}$} \\
%     \cmidrule{2-10}
%     & WER (\%) & CER (\%) & STOI (\%) & WER (\%) & CER (\%) & STOI (\%) & WER (\%) & CER (\%) & STOI (\%) \\
%     \midrule
%     sinc  & 7.13 & 4.09 & 99.76 & 12.39 & 8.09 & 89.91 & 42.90 & 32.56 & 79.04 \\
%     \midrule
%     TFiLM \cite{birnbaum2019temporal} & 7.22 & 4.20 & 99.24 & 15.16 & 10.51 & 91.27 & 55.97 & 42.59 & 80.23 \\
%     AFiLM \cite{rakotonirina2021self} & 7.13 & 4.08 & 98.54 & 14.30 & 9.29 & 90.51 & 59.95 & 45.69 & 76.83 \\
%     NVSR \cite{liu2022neural} & 7.47 & 4.19 & 98.84 & 15.91 & 10.55 & 92.04 & 53.32 & 41.10 & 82.38 \\
%     AERO \cite{mandel2023aero} & 7.17 & 4.08 & 99.38 & 11.75 & 7.39 & 93.74 & - & - & - \\
%     \textbf{AP-BWE} & \textbf{7.03} & \textbf{3.99} & \textbf{99.77} & \textbf{9.38} & \textbf{5.64} & \textbf{94.75} & \textbf{30.22} & \textbf{22.13} & \textbf{87.00} \\
%     \midrule
%     Ground Truth & 6.13  & 3.32  & 100.00 & 6.13  & 3.32  & 100.00 & 6.13  & 3.32  & 100.00 \\
%     \bottomrule
%   \end{tabular}}
% \end{table*}

\subsubsection{Evaluation on Speech Quality}
\begin{figure*}[htbp!]
  \centering
  \includegraphics[width=\linewidth]{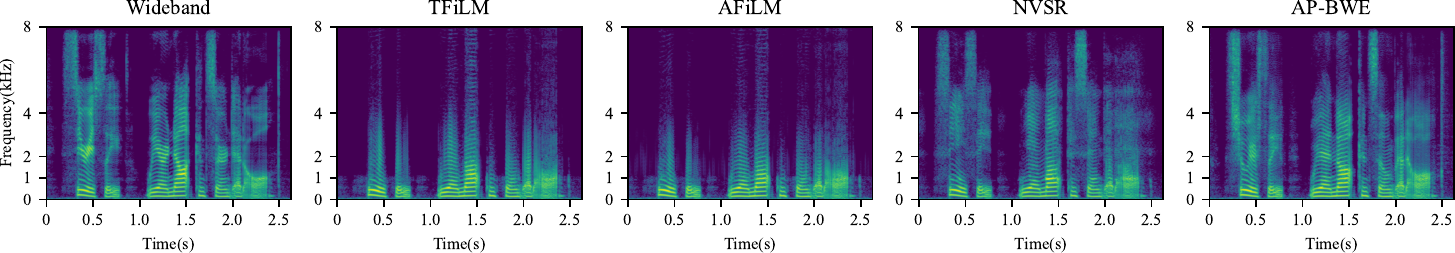}
  \caption{Spectrogram visualization of the original wideband 16 kHz speech waveform and speech waveforms extended by baseline methods and our proposed AP-BWE from the source sampling rate of 2 kHz.}
  \label{fig: specs_2kto16k}
\end{figure*}

\begin{itemize}[left=0em]
\item {}{\textbf{Objective Evaluation}}:
As depicted in Table~\ref{tab: results_16k}, our proposed AP-BWE achieved the best performance in speech quality with all kinds of source sampling rates.
Compared to sinc filter interpolation, our proposed AP-BWE exhibited significant improvements of 61.7\%, 67.5\%, and 68.6\% in terms of LSD as well as 8.5\%, 22.2\%, and 37.7\% in terms ViSQOL for source sampling rates of 8 kHz, 4 kHz, and 2 kHz, respectively.
With the narrowing of source speech bandwidth, the performance advantage of our proposed AP-BWE became more pronounced, indicating the powerful BWE capability of our model.
% and outperformed NVSR by 12.7\% and 4.2\% at the source sampling rate of 8 kHz.
In general, waveform-based methods (TFiLM and AFiLM) performed less effectively than spectrum-based methods (NVSR, AERO, and our proposed AP-BWE), indicating the importance of capturing time-frequency domain characteristics for the BWE task.
Within spectrum-based methods, NVSR, relying on high-frequency mel-spectrogram prediction and vocoder-based waveform reconstruction, demonstrated advantages in the LSD metric assessing the extended amplitude.
However, the vocoder-based phase recovery was not as effective as the complex spectrum-based approach, so it lagged behind AERO in the ViSQOL metric assessing overall speech quality.
Compared to AERO, our AP-BWE, benefiting from explicit amplitude and phase optimizations, successfully avoided the compensation effects between amplitude and phase and consequently achieved better performance in both spectral and waveform-based metrics.
Worth noting is that the unified AP-BWE* model exhibited only a slight decrease in performance compared to AP-BWE, and even achieved the highest ViSQOL score at the source sampling rate of 2 kHz.
This indicated that our model exhibited strong adaptability to the source sampling rate.

The key distinction between our approach and others lay in our implementation of explicit high-frequency phase extension.
As illustrated in Table~\ref{tab: results_pha_16k}, our proposed AP-BWE consistently outperformed other baselines across various source sampling rates, demonstrating superior performance in terms of instantaneous phase error and phase continuity along both time and frequency axes.
For the AP-BWE*, only slight decreases in the AWPD metrics were observed when the source sampling rate was 8 kHz. 
Under other source sampling rate conditions, the metrics were the same as AP-BWE, indicating the robustness of our unified model on phase.
Remarkably, for other baseline methods, some of their AWPD metrics exhibited degradation compared to those of the source sinc-interpolated waveforms. 
This suggests a limitation in the effective utilization of low-frequency phase information during the speech BWE process by these baseline methods. 
Moreover, all methods here directly generated waveforms without substituting the original low-frequency components, so their low-frequency phase might be partially compromised, leading to a significant impact on the quality of the extended speech. 
This observation underscored the critical importance of precise phase prediction and optimization in the context of BWE tasks, further emphasizing the advantage of our approach.

\begin{table}[t!]\small
  \caption{MOS Tests Results for BWE Methods with Source Sampling Rate of 2 kHz and Target Sampling Rate of 16 kHz}
  \label{tab: mos_2kto16k}
  \centering
  \begin{tabular}{lc}
    \toprule
    Methods & MOS (CI) \\
    \midrule
    sinc                              & 3.34 ($\pm$ 0.09) \\
    \midrule
    TFiLM \cite{birnbaum2019temporal} & 3.41 ($\pm$ 0.09) \\
    AFiLM \cite{rakotonirina2021self} & 3.50 ($\pm$ 0.08) \\
    NVSR \cite{liu2022neural}         & 3.68 ($\pm$ 0.08) \\
    \textbf{AP-BWE}                   & \textbf{3.93 ($\pm$ 0.07)} \\
    \midrule
    Ground Truth                      & 4.01 ($\pm$ 0.06) \\
    \bottomrule
  \end{tabular}
\end{table}

\item {}{\textbf{Subjective Evaluation}}:
To compare the BWE capabilities of our proposed AP-BWE with those of other baseline models, we conducted MOS tests on natural wideband 16 kHz speech waveforms, as well as on speech waveforms extended by AP-BWE and other baseline methods at a source sampling rate of 2 kHz. 
The subjective experimental results are presented in Table~\ref{tab: mos_2kto16k}.
For a more intuitive comparison, we visualized the spectrograms of these speech waveforms, as illustrated in Fig.\ref{fig: specs_2kto16k}. 
According to the MOS results, our proposed AP-BWE outperformed other baseline models very significantly in terms of subjective quality ($p < 0.01$).
% Overall, our proposed AP-BWE models achieved the highest MOS scores compared to the baseline models, coming close to the scores of the natural wideband speech. 
% The MOS scores for TFiLM and AFiLM were still comparable to those of sinc filter interpolation, indicating their limited ability to model high-frequency components, particularly in high-frequency unvoiced segments, as shown in Fig.\ref{fig: specs_16k}.
The MOS of TFiLM and AFiLM showed a slight improvement to that of the sinc filter interpolation, demonstrating their insufficient modeling capability for high-frequency components, particularly in the case of high-frequency unvoiced segments as shown in the in Fig.~\ref{fig: specs_2kto16k}.
NVSR achieved a decent MOS score compared to TFiLM and AFiLM but still lagged behind our proposed AP-BWE.
We can observe that, compared to the spectrograms of natural wideband speech and AP-BWE-extended speech, the NVSR-extended speech spectrogram exhibited relatively low energy in both high-frequency unvoiced segments (e.g., 0.2 $\sim$ 0.3s) and low-frequency harmonics (e.g., 1.1 $\sim$ 1.5s). 
As a result, the speech signals extended by NVSR would sound duller, negatively impacting its perceived speech quality.
In contrast, our proposed AP-BWE effectively extended more robust harmonic structures, demonstrating their strong modeling capabilities and highlighting the effectiveness of explicit predictions of amplitude and phase spectra.
% Furthermore, the unified AP-BWE* showed only a 0.04-point decrease in MOS compared to individually trained AP-BWE, further confirming the strong generalization capability of our model across different source sampling rates.

% In general, TFiLM and AFiLM exhibited insufficient modeling capability for high-frequency components compared to NVSR and our proposed AP-BWE, particularly in the case of high-frequency unvoiced segments.
% Compared to NVSR, we can observe from the spectrogram that our result had higher energy in both high-frequency unvoiced segments (e.g., 0.2 $\sim$ 0.3s) and low-frequency harmonics (e.g., 1.1 $\sim$ 1.5s). 
% As a result, the speech signals extended by our proposed AP-BWE would sound brighter, contributing to a better BWE effect.
% Compared to the natural wideband speech, even in cases where the input bandwidth was as narrow as 1 kHz, our model was still capable of extending a relatively complete harmonic structure. 
% This highlighted the robust modeling capability of our proposed AP-BWE and the effectiveness of explicit predictions of amplitude and phase spectra.

\end{itemize}

% \begin{figure*}[htbp!]
% \centering
% \subfloat[2 kHz $\rightarrow$ 16 kHz]{\includegraphics[width=\linewidth]{Figs/specs_16k.pdf}
% \label{fig: specs_16k}}
% \hfill
% \subfloat[8 kHz $\rightarrow$ 48 kHz]{\includegraphics[width=\linewidth]{Figs/specs_48k.pdf}
% \label{fig: specs_48k}}
% \caption{Spectrogram visualization of the original wideband speech waveform and speech waveforms extended by baseline methods and our proposed AP-BWE.}
% \label{fig: specs}
% \end{figure*}

% \begin{table}[t!]
%   \caption{Experimental Results of the MOS Tests under Two Target Sampling Rates with Highest Extension Ratios}
%   \label{tab: mos}
%   \centering
%   % \renewcommand\arraystretch{1.2}
%   \resizebox{\linewidth}{!}{
%   \begin{tabular}{lc|lc}
%     \toprule
%     \multicolumn{2}{c|}{$\mathrm{2 kHz \rightarrow 16 kHz}$} & \multicolumn{2}{c}{$\mathrm{8 kHz \rightarrow 48 kHz}$} \\
%     \midrule
%     Methods & MOS (CI) & Methods & MOS (CI) \\
%     \midrule
%     sinc & 3.34 ($\pm$ 0.09) & sinc & 3.69 ($\pm$ 0.07) \\
%     \midrule
%     TFiLM \cite{birnbaum2019temporal} & 3.41 ($\pm$ 0.09) & NU-Wave2 \cite{han2022nu} & 3.75 ($\pm$ 0.08) \\
%     AFiLM \cite{rakotonirina2021self} & 3.50 ($\pm$ 0.08) & UDM+ \cite{yu2023conditioning} & 3.98 ($\pm$ 0.06)\\
%     NVSR \cite{liu2022neural}         & 3.68 ($\pm$ 0.08) & mdctGAN \cite{shuai2023mdct} & 4.00 ($\pm$ 0.07) \\
%     AP-BWE*                           & 3.89 ($\pm$ 0.07) & AP-BWE* & 4.06 ($\pm$ 0.06)\\
%     \textbf{AP-BWE}                   & \textbf{3.93 ($\pm$ 0.07)} & \textbf{AP-BWE} & \textbf{4.11 ($\pm$ 0.06)} \\
%     \midrule
%     Ground Truth                      & 4.01 ($\pm$ 0.06) & Ground Truth & 4.17 ($\pm$ 0.06)\\
%     \bottomrule
%   \end{tabular}}
% \end{table}

\subsubsection{Evaluation on Generation Efficiency}

We respectively evaluated the generation efficiency of our proposed AP-BWE as well as other baseline methods as outlined in Table~\ref{tab: results_16k}.
Considering the inference speed, since NVSR divided the BWE process into the mel-spectrogram extension stage and vocoder synthesis stage, it lagged far behind other end-to-end methods.
For TFiLM and AFiLM, since they both operated on the waveform level and utilized RNNs or self-attention to capture long-term dependencies, their inference speeds were consequently constrained.
For AERO, although it and our proposed AP-BWE both operated on the spectral level, the utilization of transformer \cite{vaswani2017attention} blocks in multiple layers severely slowed down its inference speed.
Nevertheless, our AP-BWE model, based on fully convolutional networks and all-frame-level operations, has achieved an astonishingly high-speed waveform generation (29.61 times real-time speed on CPU and 382.56 times on GPU), far surpassing other baseline methods.
Considering the models' computational complexity, the FLOPs of AP-BWE were at least five times smaller than those of the baseline models, further demonstrating the advantage of our proposed model in generation efficiency.

\begin{table*}[ht!]\normalsize
  \caption{Experimental Results in Intelligibility for BWE methods Evaluated on the VCTK Dataset with Target Sampling Rate of 16 kHz}
  \label{tab: results_asr}
  \centering
  \resizebox{0.9\textwidth}{!}{
  \begin{tabular}{l|ccc|ccc|ccc}
    \toprule
    \multirow{2}{*}{Method} & \multicolumn{3}{c|}{$\mathrm{8 kHz \rightarrow 16 kHz}$} & \multicolumn{3}{c|}{$\mathrm{4 kHz \rightarrow 16 kHz}$} & \multicolumn{3}{c}{$\mathrm{2 kHz \rightarrow 16 kHz}$} \\
    \cmidrule{2-10}
    & WER (\%) & CER (\%) & STOI (\%) & WER (\%) & CER (\%) & STOI (\%) & WER (\%) & CER (\%) & STOI (\%) \\
    \midrule
    sinc  & \textbf{3.67} & \textbf{1.67} & 99.76 & 11.45 & 7.08 & 89.91 & 47.43 & 33.56 & 79.04 \\
    \midrule
    TFiLM \cite{birnbaum2019temporal} & 3.69 & 1.69 & 99.24 & 11.32 & 7.24 & 91.27 & 45.95 & 33.58 & 80.23 \\
    AFiLM \cite{rakotonirina2021self} & \textbf{3.67} & \textbf{1.67} & 98.54 & 9.28 & 5.53 & 90.51 & 45.16 & 33.01 & 76.83 \\
    NVSR \cite{liu2022neural}         & 4.38 & 2.02 & 98.84 & 13.56 & 8.51 & 92.04 & 59.53 & 44.43 & 82.38 \\
    AERO \cite{mandel2023aero}        & 3.97 & 1.84 & 99.38 & 9.78 & 5.51 & 93.74 & - & - & - \\
    \textbf{AP-BWE}                   & 3.72 & \textbf{1.67} & \textbf{99.77} & \textbf{6.69} & \textbf{3.54} & \textbf{94.75} & \textbf{36.69} & \textbf{25.61} & \textbf{87.00} \\
    \midrule
    Ground Truth                      & 3.07 & 1.26 & 100.00 & 3.07 & 1.26 & 100.00 & 3.07 & 1.26 & 100.00 \\
    \bottomrule
  \end{tabular}}
\end{table*}

\subsubsection{Evaluation on Speech Intelligibility}

As shown in Table~\ref{tab: results_asr}, it is obvious that our proposed AP-BWE exhibited a remarkable improvement in terms of intelligibility metrics compared to baseline models.
Under the condition of extending from 8 kHz to 16 kHz, the performance of the sinc filter interpolation was already very close to the Ground Truth. 
Our proposed AP-BWE and other baseline models struggled to further improve WER and CER on top of the waveform interpolated by the sinc filter, suggesting that the ASR model focused on information from frequencies below 4 kHz for transcription.
When the source sampling rate was further reduced to 4 kHz and 2 kHz, all the baseline models showed slight improvements in WER and CER compared to sinc filter interpolation, except for NVSR. 
The decline in NVSR’s performance in WER and CER was due to its use of a vocoder to restore the waveform, which made the low-frequency components unnatural, but its STOI metric was still improved.
Overall, these baseline models demonstrated limited extension capabilities under the extremely high extension ratio.
However, our proposed AP-BWE significantly improved WER, CER, and STOI by 41.57\%, 50.00\%, 5.38\% at the 4 kHz source sampling rate, and by 22.64\%, 23.69\%, and 10.07\% at the 2 kHz source sampling rate, compared to sinc filter interpolation.
This indicated that benefiting from our precise phase prediction, our model possessed strong harmonic restoration capabilities, reconstructing the key information of vowels and consonants as well as significantly enhancing the intelligibility of the extended speech.

\begin{table*}[htbp!]\small
  \caption{Experimental Results in Speech Quality (LSD and ViSQOL) and Generation Efficiency (RTF and FLOPs) for BWE Methods Evaluated on the VCTK Dataset with Target Sampling Rate of 48 kHz, Where in RTF ($a\times$) Representing $a$ Times Real-Time}
  \label{tab: results_48k}
  \centering
  \resizebox{\textwidth}{!}{
  \begin{tabular}{l|cc|cc|cc|cc|cc|c}
    \toprule
    \multirow{2}{*}{Method} & \multicolumn{2}{c|}{$\mathrm{24 kHz \rightarrow 48 kHz}$} & \multicolumn{2}{c|}{$\mathrm{16 kHz \rightarrow 48 kHz}$} & \multicolumn{2}{c|}{$\mathrm{12 kHz \rightarrow 48 kHz}$} & \multicolumn{2}{c|}{$\mathrm{8 kHz \rightarrow 48 kHz}$} & \multirow{2}{*}{$\mathrm{RTF (CPU)}$} & \multirow{2}{*}{$\mathrm{RTF (GPU)}$} & \multirow{2}{*}{$\mathrm{FLOPs}$} \\ 
    \cmidrule{2-9}
    & $\mathrm{LSD}$ & $\mathrm{ViSQOL}$ & $\mathrm{LSD}$ & $\mathrm{ViSQOL}$ & $\mathrm{LSD}$ & $\mathrm{ViSQOL}$ & $\mathrm{LSD}$ & $\mathrm{ViSQOL}$ &  &  & \\
    \midrule
    sinc & 2.17 & 2.99 & 2.57 & 2.26 & 2.75 & 2.09 & 2.94 & 2.07 & - & - & - \\
    \midrule
    NU-Wave \cite{lee2021nu} & 0.85 & 3.18 & 0.99 & 2.36 & -    & -    & -    & -    & 95.57 (0.01$\times$) & 0.5018 (1.99$\times$) & 4039.13G \\
    NU-Wave2 \cite{han2022nu} & 0.72 & 3.74 & 0.86 & 3.00 & 0.94 & 2.75 & 1.09 & 2.48 & 92.58 (0.01$\times$) & 0.5195 (1.92$\times$) & 1385.27G \\
    UDM+ \cite{yu2023conditioning} & 0.64 & 4.02 & 0.79 & 3.35 & 0.88 & 3.08 & 1.03 & 2.81 & 74.03 (0.01$\times$) & 0.8335 (1.20$\times$) & 2369.50G \\
    % mdctGAN \cite{shuai2023mdct} & 0.72 & 3.58 & 0.82 & 3.15 & 0.90 & 2.96 & 0.93 & 2.95 & 0.2461 (4.06$\times$) & 0.0129 (77.80$\times$)\\
    mdctGAN \cite{shuai2023mdct} & 0.71 & 3.69 & 0.83 & 3.27 & 0.85 & 3.12 & 0.93 & 3.03 & 0.2461 (4.06$\times$) & 0.0129 (77.80$\times$) & 103.38G \\
    \midrule
    AP-BWE*   & 0.62 & 4.17 & \textbf{0.72} & 3.63 & 0.79 & \textbf{3.46} & 0.85 & 3.32 & \multirow{2}{*}{\textbf{0.0551 (18.14$\times$)}} & \multirow{2}{*}{\textbf{0.0034 (292.28$\times$)}} & \multirow{2}{*}{\textbf{17.87G}} \\
    \textbf{AP-BWE} & \textbf{0.61} & \textbf{4.25} & \textbf{0.72} & \textbf{3.70} & \textbf{0.78} & \textbf{3.46} & \textbf{0.84} & \textbf{3.35} & & \\
    \bottomrule
  \end{tabular}}
\end{table*}
\begin{table*}[htbp!]\Large
  \caption{Experimental Results for the Band-wise Analysis with Source Sampling Rate of 8 kHz and Target Sampling Rate of 48 kHz}
  \label{tab: band-wise analysis}
  \centering
  \resizebox{\textwidth}{!}{
  \begin{tabular}{l|cccc|cccc|cccc}
    \toprule
    \multirow{2}{*}{Method} & \multicolumn{4}{c|}{$\mathrm{4 kHz \sim 8 kHz}$} & \multicolumn{4}{c|}{$\mathrm{8 kHz \sim 12 kHz}$} & \multicolumn{4}{c}{$\mathrm{12 kHz \sim 24 kHz}$} \\
    \cmidrule{2-13}
    & $\mathrm{LSD}$ & $\mathrm{AWPD}_{IP}$ & $\mathrm{AWPD}_{GD}$ & $\mathrm{AWPD}_{IAF}$ & $\mathrm{LSD}$ & $\mathrm{AWPD}_{IP}$ & $\mathrm{AWPD}_{GD}$ & $\mathrm{AWPD}_{IAF} $ & $\mathrm{LSD}$ & $\mathrm{AWPD}_{IP}$ & $\mathrm{AWPD}_{GD}$ & $\mathrm{AWPD}_{IAF}$ \\
    \midrule
    NU-Wave2 \cite{han2022nu} & 1.35 & 1.81 & 1.48 & 1.47 & 1.24 & 1.81 & 1.48 & 1.47 & 1.09 & 1.82 & 1.47 & 1.46 \\
    UDM+ \cite{yu2023conditioning} & 1.21 & 1.80 & 1.45 & 1.46 & 1.26 & 1.81 & 1.46 & \textbf{1.46} & 1.03 & 1.82 & 1.46 & 1.46 \\
    % mdctGAN \cite{shuai2023mdct} & 1.05 & 1.80 & 1.46 & 1.46 & 1.03 & 1.81 & 1.46 & \textbf{1.46} & 0.98 & 1.82 & 1.46 & 1.46 \\
    mdctGAN \cite{shuai2023mdct} & 1.11 & 1.80 & 1.45 & 1.46 & 1.07 & 1.81 & 1.46 & 1.47 & 0.93 & 1.82 & 1.46 & 1.46 \\
    \textbf{AP-BWE} & \textbf{0.98} & \textbf{1.75} & \textbf{1.41} & \textbf{1.44} & \textbf{0.98} & 1.81 & \textbf{1.44} & \textbf{1.46} & \textbf{0.86} & 1.82 & \textbf{1.45} & 1.46 \\
    \bottomrule
  \end{tabular}}
\end{table*}

\subsection{BWE Experiments Targeting 48 kHz}
\subsubsection{Baseline Methods}
For BWE targeting a 48 kHz sampling rate, the sinc filter interpolation was still used as the low-bound method.
We subsequently compared our proposed AP-BWE with three diffusion-based methods (NU-Wave \cite{lee2021nu}, NU-Wave 2 \cite{han2022nu}, and UDM+ \cite{yu2023conditioning}) and an MDCT-spectrum-based method (mdctGAN \cite{shuai2023mdct}).
For NU-Wave, we used the community-contributed checkpoints from their official implementation \footnote{\href{https://github.com/maum-ai/nuwave}{https://github.com/maum-ai/nuwave}.}.
Notably, NU-Wave did not conduct the experiment at source sampling rates of 8 kHz and 12 kHz, so we excluded these results from our analysis.
For NU-Wave 2 and UDM+, we used the reproduced NU-Wave 2 checkpoint and official UDM+ checkpoint \footnote{\href{https://github.com/yoyololicon/diffwave-sr}{https://github.com/yoyololicon/diffwave-sr}.}.
It is worth noting that in the original paper of mdctGAN \cite{shuai2023mdct}, the mdctGAN model was trained on the combination of the VCTK training set and the HiFi-TTS dataset, and tested on the VCTK test set. 
% Here, we employed the checkpoints from its official implementation \footnote{\href{https://github.com/neoncloud/mdctGAN}{https://github.com/neoncloud/mdctGAN}.}.
Here, for a fair comparison, we re-trained all the mdctGAN models solely on the VCTK training set following its official implementation \footnote{\href{https://github.com/neoncloud/mdctGAN}{https://github.com/neoncloud/mdctGAN}.}.
In addition, the AP-BWE* was trained using randomly selected sampling rates from 8 kHz, 12 kHz, 16 kHz, and 24 kHz to handle inputs of various resolutions.

\subsubsection{Evaluation on Speech Quality}
\begin{itemize}[left=0em]
\item {}{\textbf{Objective Evaluation}}:

As depicted in Table~\ref{tab: results_48k}, for the high-sampling-rate waveform generation at 48 kHz, our proposed AP-BWE still achieved the SOTA performance in objective metrics, irrespective of the source sampling rates.
In general, compared to the baseline models, our approach exhibited a notably significant improvement in ViSQOL, particularly under lower extension ratios, which underscored the substantial impact of precise phase prediction on speech quality.
For diffusion-based methods, since both NU-Wave2 and UDM+ implemented a single model for extension across different source sampling rates, we compared our unified AP-BWE* model with them.
Compared to NU-Wave 2 and UDM+, our AP-BWE* model exhibited growing superiority in LSD as the source sampling rate decreased. 
This suggested that diffusion-based methods, operating at the waveform level, struggled to effectively recover spectral information in scenarios with restricted bandwidth.
Although both mdctGAN and our proposed AP-BWE were spectrum-based methods, AP-BWE significantly outperformed it across all source sampling rates.
Especially in terms of the overall speech quality, our proposed AP-BWE surpassed mdctGAN by 15.2\%, 13.1\%, 10.9\%, and 10.6\% in ViSQOL at source sampling rates of 24 kHz, 16 kHz, 12 kHz, and 8 kHz, respectively. 
This suggested that STFT spectra were more suitable for waveform generation tasks compared to MDCT spectra.
Additionally, similar to the results at 16 kHz, our unified AP-BWE* model demonstrated competence across different sampling rate inputs, with only a slight decline in quality compared to AP-BWE, reaffirming the adaptability of our approach to source sampling rates.

Unlike the strong harmonic structure observed in the high-frequency components of 16 kHz waveforms, the high-frequency portion of 48 kHz waveforms exhibited more randomness. 
In our preliminary experiments, we observed the phase metrics of extended 48 kHz waveforms show minimal variation between systems, especially in scenarios with relatively higher source sampling rates.
Therefore, with the source sampling rate of 8 kHz, we calculated LSD and AWPD separately for different frequency bands of the extended 48 kHz waveform to assess the performance of these models within different frequency ranges, and the evaluation results are depicted in Table~\ref{tab: band-wise analysis}.
For the LSD metric, our AP-BWE outperformed other baseline models within each frequency band.
Regarding the AWPD metrics, we only exhibited an advantage in the 4 kHz $\sim$ 8 kHz frequency band, while the differences between systems were minimal in the 8 kHz $\sim$ 12 kHz and 12 kHz $\sim$ 24 kHz frequency bands.
This indicated that our proposed AP-BWE, benefiting from explicit phase prediction, was capable of effectively recovering the harmonic structure in the waveform, thereby significantly improving the speech quality in the mid-to-low-frequency range.
For high-frequency phases, due to their strong randomness, the current methods exhibited comparable predictive capabilities.

\begin{figure*}[htbp!]
  \centering
  \includegraphics[width=\linewidth]{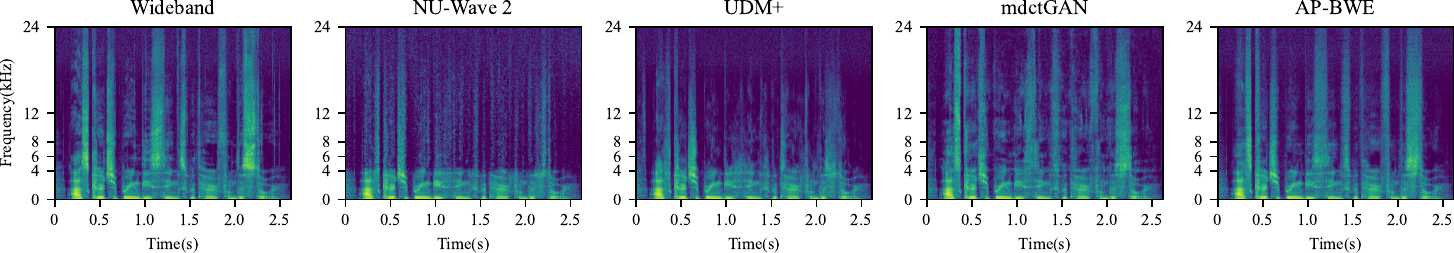}
  \caption{Spectrogram visualization of the original wideband 48 kHz speech waveform and speech waveforms extended by baseline methods and our proposed AP-BWE from the source sampling rate of 8 kHz.}
  \label{fig: specs_8kto48k}
\end{figure*}

\begin{table}[t!]\small
  \caption{MOS Tests Results for BWE Methods with Source Sampling Rate of 8 kHz and Target Sampling Rate of 48 kHz}
  \label{tab: mos_8kto48k}
  \centering
  \begin{tabular}{lc}
    \toprule
    Methods & MOS (CI) \\
    \midrule
    sinc                              & 3.69 ($\pm$ 0.07) \\
    \midrule
    NU-Wave2 \cite{han2022nu}         & 3.75 ($\pm$ 0.08) \\
    UDM+ \cite{yu2023conditioning}    & 3.98 ($\pm$ 0.06) \\
    mdctGAN \cite{shuai2023mdct}      & 4.00 ($\pm$ 0.07) \\
    \textbf{AP-BWE}                   & \textbf{4.11 ($\pm$ 0.06)} \\
    \midrule
    Ground Truth                      & 4.17 ($\pm$ 0.06) \\
    \bottomrule
  \end{tabular}
\end{table}

\item {}{\textbf{Subjective Evaluation}}:
As shown in the right half of Table~\ref{tab: mos_8kto48k}, we conducted MOS tests on the wideband 48 kHz speech waveform, along with speech waveforms extended by AP-BWE and other baseline methods at a source sampling rate of 8 kHz. 
We also visualized the corresponding spectrograms, as illustrated in Fig.~\ref{fig: specs_8kto48k}.
In this configuration, the initial 4 kHz bandwidth already contained the full fundamental frequency and most harmonic structures, resulting in less pronounced subjective listening differences between the speech extended by different models.
Nevertheless, the MOS results demonstrated that our proposed AP-BWE still showed substantial advantages in subjective quality over other baseline models ($p < 0.05$).
Firstly, NU-Wave2 scored very significantly lower in MOS compared to our proposed AP-BWE ($p < 0.01$), showing only a slight improvement over sinc filter interpolation, with spectrogram analysis revealing poor recovery of mid-to-high frequency components.
% Both UDM+ and mdctGAN achieved significantly higher MOS than NU-Wave2.
UDM+ performed well in recovering mid-frequency components of speech, but it seemed to struggle with restoring higher-frequency components, particularly with low energy in the unvoiced segments, resulting in extended speech that sounded less bright.
Consequently, the subjective quality of UDM+ remained significantly lower than that of our proposed AP-BWE ($p = 0.020$).
This finding aligned with the results obtained at the target sampling rate of 16 kHz, suggesting a potential limitation in modeling high-frequency unvoiced segments for waveform-based methods. 
The mdctGAN achieved the optimal MOS among the baseline methods, with its corresponding spectrogram displaying brighter and more complete structures.
However, the high-frequency components of the spectrogram exhibited higher randomness and poorer continuity, resulting in a less stable auditory perception. 
In contrast, our proposed AP-BWE demonstrated a more robust restoration capability for the high-frequency components, especially in the unvoiced segments, giving it a substantial advantage in subjective speech quality over mdctGAN ($p = 0.026$).
% Therefore, the AP-BWE models held the highest MOS and even approaching that of the wideband 48 kHz speech.
% Spectrogram-based models, including mdctGAN and our proposed AP-BWE models, achieved significantly higher MOS scores, with our models holding a slight advantage.
% It can also be observed that the 
% In contrast, our proposed AP-BWE demonstrated robust restoration capabilities for the high-frequency components, especially in the unvoiced segments. 
While there were still differences in the high-frequency components of the voiced segments compared to natural wideband speech, these distinctions had minimal impact on the perceptual quality of the speech.
Therefore, AP-BWE achieved a MOS close to that of natural wideband speech.
% The spectrograms revealed that NU-Wave2 and UDM+ displayed relatively weaker recovery capabilities for high-frequency components, particularly with low energy in the unvoiced segments, resulting in extended speech that sounded less bright.

\end{itemize}

\subsubsection{Evaluation on Generation Efficiency}
Considering the inference speed, as depicted in Table~\ref{tab: results_48k}, our proposed AP-BWE remained capable of efficient 48 kHz speech waveform production at a speed of 18.14 times real-time on CPU and 292.28 times real-time on GPU.
For diffusion-based methods (i.e., NU-Wave, NU-Wave2, and UDM+), their generation efficiency took a significant hit as they require multiple time steps in the reverse process to continuously denoise and recover the extended waveform from latent variables.
Remarkably, our AP-BWE achieved a speedup over them by approximately 1000 times on CPU and 100 times on GPU, respectively.
Despite both mdctGAN and our proposed AP-BWE operating at the spectral level, the generation speed of mdctGAN was still constrained by its two-dimensional convolution and transformer-based structure. 
Consequently, our AP-BWE which is fully based on one-dimensional convolutions enabled an approximately fourfold acceleration compared to it.
Compared to running on a GPU, our model exhibited a more significant efficiency improvement on the CPU. 
This indicated that our model could efficiently generate high-sampling-rate samples even without the parallel acceleration support of GPUs, making it more suitable for applications in scenarios with limited computational resources.
Considering the models' computational complexity, the FLOPs of diffusion models are heavily constrained by their reverse steps (82.78G $\times$ 50 steps for NU-Wave, 27.71G $\times$ 50 steps for NU-Wave2, and 47.39G $\times$ 50 steps for UDM+). 
However, even with just one step generation, the FLOPs of our proposed AP-BWE were still smaller than theirs, further demonstrating the superiority of our proposed AP-BWE in terms of generation efficiency.
Comparing Table~\ref{tab: results_16k} and Table~\ref{tab: results_48k}, it can be observed that for generating speech waveforms of the same duration, the inference speed of our model at 48 kHz sampling rate was relatively lower, while the computational complexity was higher compared to the 16 kHz sampling rate.
This was attributed to the fact that our model under both sampling rate configurations utilized the same STFT settings, resulting in different frame numbers processed by the model.

\begin{figure*}[htbp!]
  \centering
  \includegraphics[width=\linewidth]{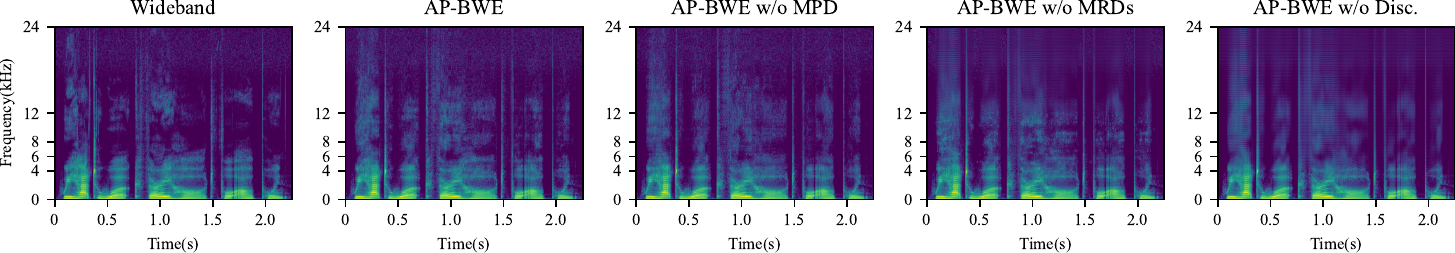}
  \caption{Spectrogram visualization of the original wideband speech waveform and speech waveforms extended by the ablation models of our proposed AP-BWE with a source sampling rate of 8 kHz and target sampling rate of 48 kHz. ``AP-BWE w/o MRDs" represents the ablation of both MRAD and MRPD, while ``AP-BWE w/o Disc." denotes the ablation of all discriminators.}
  \label{fig: Ablation}
\end{figure*}

\subsection{Analysis and Discussion}
\subsubsection{Ablation Studies}
We implemented ablation studies on discriminators and between-stream connections to investigate the roles of each discriminator and the effects of interactions between amplitude stream and phase stream.
All the experiments were conducted with the source sampling rate of 8 kHz and target sampling rate of 48 kHz, and the experimental results are depicted in Table~\ref{tab: ablation}.
Due to the minimal phase differences in the high-frequency components for the BWE targeting 48 kHz, we calculated the AWPD metric only in the frequency band of 4 kHz to 8 kHz while calculating the LSD metric on the whole frequency band.
We further visualized the spectrograms of the natural wideband 48 kHz speech waveform and speech waveforms generated by the ablation models of AP-BWE on discriminator, as illustrated in Fig.~\ref{fig: Ablation}.

\begin{table}[t!]\LARGE
  \caption{Experimental Results for the Ablation Studies with Source Sampling Rate of 8 kHz and Target Sampling Rate of 48 kHz}
  \label{tab: ablation}
  \centering
  \resizebox{\linewidth}{!}{
  \begin{tabular}{cc|cccccc}
    \toprule
    \multicolumn{7}{c}{\textbf{Ablation on Discriminators}} \\
    \midrule
    MPD & MRAD / MRPD & $\mathrm{LSD}$ & $\mathrm{AWPD}_{IP}$ & $\mathrm{AWPD}_{GD}$ & $\mathrm{AWPD}_{IAF}$ & $\mathrm{ViSQOL}$ \\
    \midrule
    \Checkmark & \Checkmark/\Checkmark       & \textbf{0.84} & 1.75 & \textbf{1.41} & \textbf{1.44} & \textbf{3.35} \\
    \XSolidBrush & \Checkmark/\Checkmark     & 0.85 & 1.76 & 1.42 & \textbf{1.44} & 3.29 \\
    \Checkmark & \XSolidBrush/\Checkmark     & 0.86 & 1.75 & \textbf{1.41} & \textbf{1.44} & 3.26 \\
    \Checkmark & \Checkmark/\XSolidBrush     & 0.85 & 1.76 & 1.42 & \textbf{1.44} & 3.31 \\
    \Checkmark & \XSolidBrush/\XSolidBrush   & 0.88 & 1.75 & \textbf{1.41} & \textbf{1.44} & 3.26 \\
    % \XSolidBrush & \Checkmark/\XSolidBrush   & 0.85 & 1.75 & \textbf{1.41} & \textbf{1.44} & 3.27 \\
    % \XSolidBrush & \XSolidBrush/\Checkmark   & 0.86 & 1.75 & \textbf{1.41} & \textbf{1.44} & 3.26 \\
    \XSolidBrush & \XSolidBrush/\XSolidBrush & 1.50 & \textbf{1.74} & 1.42 & 1.50 & 3.26 \\
    \midrule
    \multicolumn{7}{c}{\textbf{Ablation on Between-Stream Connections}} \\
    \midrule
    A $\rightarrow$ P & P $\rightarrow$ A & $\mathrm{LSD}$ & $\mathrm{AWPD}_{IP}$ & $\mathrm{AWPD}_{GD}$ & $\mathrm{AWPD}_{IAF}$ & $\mathrm{ViSQOL}$ \\
    \midrule
    \Checkmark & \Checkmark     & \textbf{0.84} & \textbf{1.75} & \textbf{1.41} & \textbf{1.44} & \textbf{3.35}\\
    \XSolidBrush & \Checkmark   & 0.85 & 1.77 & 1.42 & 1.45 & 3.31 \\
    \Checkmark & \XSolidBrush   & 0.85 & 1.76 & 1.42 & \textbf{1.44} & 3.32 \\
    \XSolidBrush & \XSolidBrush & 0.86 & 1.77 & 1.42 & 1.45 & 3.32 \\
    \bottomrule
  \end{tabular}}
\end{table}

As shown in the upper half of Table~\ref{tab: ablation}, for the ablation of discriminators, all the discriminators contributed to the overall performance of our proposed AP-BWE.
We first ablated the MPD to train the AP-BWE model solely with discriminators at the spectral level.
Although there is only a slight decrease in all metrics, the spectrogram (AP-BWE w/o MPD) reveals a smearing effect along the frequency axis as shown in Fig.~\ref{fig: Ablation}, resulting in a perceptible harshness in the extended speech.
Subsequently, in our preliminary experiments, we separately ablated MRPD and MRAD. 
In both cases, the metrics showed only slight decreases, and the spectrograms appeared normal.
However, when we simultaneously ablated both of them (AP-BWE w/o MRDs), although the metrics still decreased insignificantly, a noticeable over-smoothing could be observed in the frequency band from 12 kHz to 24 kHz of the spectrogram.
This is because with the sole utilization of MPD, the minimum period of its sub-discriminator was 2, so the frequency band range it can discriminate was only from 0 to 12 kHz.
As this over-smoothing phenomenon is present in the high-frequency range, it has not had a substantial impact on the perceived quality of extended speech.
We further ablated all discriminators (AP-BWE w/o Disc.), the LSD metric experienced a significant decline, and the extended portions across the entire spectrogram exhibited severe over-smoothing, greatly compromising the quality of the extended speech.
This indicated that the training strategy of GAN was indispensable for the current AP-BWE model.

Moreover, we ablated the between-stream connections.
As shown in the last row of Table~\ref{tab: ablation}, the information interaction between the amplitude stream and the phase stream did contribute to the quality of the extended speech.
To investigate the influence of one stream on another, we selectively ablated each of the connections.
Our observations revealed that when ablating the connection from the amplitude stream to the phase stream (A $\rightarrow$ P), the AWPD metrics exhibited a deterioration compared to ablating the connection from the phase stream to the amplitude stream (P $\rightarrow$ A), and there was also a decrease in ViSQOL, indicating that amplitude information played a role in the modeling of phase.
This conclusion aligns with our previous work \cite{ai2023neural}, where the phase spectrum can be predicted from the amplitude spectrum.

\begin{table}[t!]\normalsize
  \caption{Experimental Results for the Cross-Dataset Evaluation on the Libri-TTS and HiFi-TTS Datasets}
  \label{tab: cross-dataset}
  \centering
  \resizebox{\linewidth}{!}{
  \begin{tabular}{l|ccccc}
    \toprule
    \multicolumn{6}{c}{\textbf{Libri-TTS (8 kHz $\rightarrow$ 24 kHz)}} \\
    \midrule
    Methods & $\mathrm{LSD}$ & $\mathrm{AWPD}_{IP}$ & $\mathrm{AWPD}_{GD}$ & $\mathrm{AWPD}_{IAF}$  & ViSQOL \\
    \midrule
    NU-Wave2 & 1.83 & 1.82 & 1.51 & 1.47 & 2.92 \\
    UDM+     & 1.79 & 1.82 & 1.50 & 1.46 & 2.88 \\
    % mdctGAN  & 1.28 & 1.80 & 1.44 & 1.46 & 3.38 \\
    mdctGAN  & 1.27 & 1.80 & 1.44 & 1.46 & 3.42 \\
    \textbf{AP-BWE}  & \textbf{1.22} & \textbf{1.79} & \textbf{1.40} & \textbf{1.44} & \textbf{3.44} \\
    \midrule
    \multicolumn{6}{c}{\textbf{HiFi-TTS (8 kHz $\rightarrow$ 44.1 kHz)}} \\
    \midrule
    Methods & $\mathrm{LSD}$ & $\mathrm{AWPD}_{IP}$ & $\mathrm{AWPD}_{GD}$ & $\mathrm{AWPD}_{IAF}$ & ViSQOL\\
    \midrule
    NU-Wave2 & 1.67 & 1.81 & 1.48 & 1.48 & 2.16 \\
    UDM+     & 1.56 & 1.80 & 1.47 & 1.47 & 2.17 \\
    % mdctGAN  & \textbf{1.49} & 1.80 & 1.46 & 1.47 & \textbf{2.57} \\
    mdctGAN  & 1.69 & 1.80 & 1.47 & 1.47 & 2.43 \\
    AP-BWE   & \textbf{1.49} & \textbf{1.77} & \textbf{1.42} & \textbf{1.45} & \textbf{2.51} \\
    \bottomrule
  \end{tabular}}
\end{table}

\subsubsection{Cross-Dataset Validation}
Since the speech data in a corpus is recorded in a fixed environment, models trained exclusively on a single corpus may adapt to the specific characteristics of the recording environment.
To evaluate the models' generalization abilities across different corpora, we conducted cross-dataset experiments on models trained with the source and target sampling rates of 8 kHz and 48 kHz, respectively.
We selected two high-quality datasets, namely Libri-TTS \cite{zen2019libritts} and HiFi-TTS \cite{bakhturina2021hifitts}.
The Libri-TTS dataset consists of 585 hours of speech data at a 24 kHz sampling rate.
For evaluation, we exclusively utilized the ``test-clean" set, containing 4,837 audio clips.
The HiFi-TTS dataset contains about 292 hours of speech from 10 speakers with at least 17 hours per speaker sampled at 44.1 kHz, and we also only evaluated the models on its test set which contains 1000 audio clips.
The experimental results are depicted in Table~\ref{tab: cross-dataset}, where the LSD scores were computed by downsampling all the extended speech waveforms from 48 kHz to the original sampling rates of the datasets, the AWPD metrics were calculated only in the 4kHz $\sim$ 8 kHz frequency band for a more intuitive comparison, and the ViSQOL scores were computed by upsampling all the speech waveforms to 48 kHz.

For the evaluation on the Libri-TTS dataset, as depicted in the upper half in Table~\ref{tab: cross-dataset}, our proposed AP-BWE still achieved the best performance among all the metrics.
For NU-Wave2 and UDM+, their performance on the Libri-TTS dataset was noticeably degraded compared to VCTK. This indicated a strong dependency of waveform-based methods on the training corpus, whereas spectrum-based approaches, by capturing temporal and spectral characteristics from the waveforms, exhibited adaptability to various data recording environments.
The evaluation results on the HiFi-TTS dataset are depicted in the lower half in Table~\ref{tab: cross-dataset}.
Compared to the waveform-based methods, spectrum-based methods still outperform in terms of the overall speech quality as indicated by the ViSQOL metric.
Compared to NU-Wave2 and UDM+, the advantage of mdctGAN on the HiFi-TTS test set is far less pronounced than on the Libri-TTS test set, especially in terms of the LSD metric. 
This suggests that different models exhibit varying generalization abilities across different datasets.
However, our proposed AP-BWE still shows a significant advantage over other baseline models across all metrics, further demonstrating the superior generalization ability of our model.
% Spectrum-based methods consistently outperform waveform-based methods, particularly in terms of the ViSQOL metric.
% Unsurprisingly, mdctGAN achieved the best performance since it was trained on the combination of the VCTK and HiFi-TTS datasets.
% However, our AP-BWE, trained solely on VCTK, still achieved a comparable performance and even surpassed mdctGAN in the AWPD metrics. 
% This suggests that our phase prediction exhibited sufficient generalization across different corpora.
% The overall performance of the models on the HiFi-TTS dataset was not as impressive as on Libri-TTS. 
% This discrepancy could be attributed to the fact that, despite the HiFi-TTS dataset having a sampling rate of 44.1 kHz, its effective bandwidth falls between 13 kHz and 22.05 kHz. 
% In contrast, the waveforms generated by the BWE models cover the entire bandwidth. 
% As a result, there may be mismatches in the calculation of metrics.

\section{Conclusion}
In this paper, we introduced AP-BWE, a GAN-based BWE model that can efficiently achieve high-quality wideband waveform generation.
The generator of AP-BWE performed direct recovery of high-frequency amplitude and phase information from the narrowband amplitude and phase spectra through an all-convolutional structure and all-frame-level operations, significantly enhancing generation efficiency.
Moreover, multiple discriminators applied on the time-domain waveform, amplitude spectrum, and phase spectrum noticeably elevated the overall generation quality.
The major contribution of the AP-BWE lay in the direct extension of the phase spectrum.
This allowed for the precise modeling and optimization of both the amplitude and phase spectra simultaneously, significantly enhancing the quality of the extended speech without compromising the trade-offs between the two.
Experimental results on the VCTK-0.92 dataset showcased that our proposed AP-BWE achieved SOTA performance for tasks with target sampling rates of both 16 kHz and 48 kHz. 
Spectrogram visualizations underscored the robust capability of our model in recovering high-frequency harmonic structures, effectively enhancing the intelligibility of speech signals, even in scenarios with extremely low source speech bandwidth.
% Furthermore, ablation studies verified the contributions of each discriminator to the extended speech quality and the mutual guiding effect between amplitude and phase.
% Lastly, cross-dataset experiments demonstrated the excellent generalization capability of our model across different corpora. 
In future work, our AP-BWE model can be further applied to assist generative models trained on low-sampling-rate datasets in improving their synthesized speech quality.

\bibliographystyle{IEEEtran}
\bibliography{mybib}

\end{document}